\documentclass[12pt]{iopart}
\usepackage{graphicx} 
\usepackage{color}
\definecolor{purple}{rgb}{0.5,0,0.5}

\usepackage[colorlinks=true,
            pdfstartview=FitV,
        linkcolor=purple,
        citecolor=purple,
        urlcolor=blue,
        pdftitle={Lattice Variational Approach},
        pdfpagemode=None
       ]{hyperref}
\usepackage{epsfig}

\newcommand{\sfrac}[2]{\mbox{\footnotesize $\displaystyle \frac{#1}{#2}$}} 
\newcommand{\nlsim}{\mathrel{\rlap{\lower4pt\hbox{\hskip0pt$\sim$}} 
 \raise1pt\hbox{$<$}}}           
\newcommand{\ngsim}{\mathrel{\rlap{\lower4pt\hbox{\hskip0pt$\sim$}} 
 \raise1pt\hbox{$>$}}}           

\def\deg{^\circ}
\def\gtorder{\mathrel{\raise.3ex\hbox{$>$}\mkern-14mu
 \lower0.6ex\hbox{$\sim$}}}
\def\ltorder{\mathrel{\raise.3ex\hbox{$<$}\mkern-14mu
 \lower0.6ex\hbox{$\sim$}}}

\def\mugegm{\mu_p G_E / G_M}
\def\gegm{G_E / G_M}
\def\ge{G_E}
\def\gm{G_M}
\def\mugegmp{\mu_p G_{E}^p / G_{M}^p}
\def\gegmp{G_{E}^p / G_{M}^p}
\def\gep{G_{E}^p}
\def\gmp{G_{M}^p}

\def\gegmn{G_{E}^n / G_{M}^n}
\def\gen{G_{E}^n}
\def\gmn{G_{M}^n}

\begin{document}

\topical[Nucleon electromagnetic form factors]
{Nucleon electromagnetic form factors
}

\author{J~Arrington$^1$, C~D~Roberts$^1$ and J~M~Zanotti$^2$}

\address{$^1$ Physics Division, Argonne National Laboratory,
Argonne, IL, 60439, USA}

\address{$^2$ School of Physics, University of Edinburgh, Edinburgh, EH9 3JZ, UK}


\begin{abstract}
Elastic electromagnetic nucleon form factors have long provided vital information about the structure and composition of these most basic elements of nuclear physics.  The form factors are a measurable and physical manifestation of the nature of the nucleons' constituents and the dynamics that binds them together.  Accurate form factor data obtained in recent years using modern experimental facilities has spurred a significant reevaluation of the nucleon and pictures of its structure; e.g., the role of quark orbital angular momentum, the scale at which perturbative QCD effects should become evident, the strangeness content, and meson-cloud effects.  We provide a succinct survey of the experimental studies and theoretical interpretation of nucleon electromagnetic form factors.  
\end{abstract}




\section{Introduction}
It might be said that the impact of information on nucleon electromagnetic form factors was first felt in 1933 when Otto Stern measured the proton's magnetic moment:
\begin{equation}
\label{stern}
\mu_p = (1 + \mbox{\underline{1.79}}) \frac{e}{2 M}\,.
\end{equation}
The deviation from one expressed by the underlined term within the parentheses announced that the proton is not a point particle.  Were the proton point-like, it would be described by Dirac's theory of relativistic fermions and hence have a magnetic moment, $\mu_p=\mu_D = e/(2M)$.

The impact continues to the present day, with modern, high-luminosity experimental facilities that employ large momentum transfer reactions providing remarkable and intriguing new information on nucleon structure \cite{gao03,leeburkert,HAPPEx}.  For examples one need only look so far as the discrepancy between the ratio of electromagnetic proton form factors extracted via Rosenbluth separation and that inferred from polarisation transfer \cite{punjabi05,gayou02,arrington04a,qattan05} and the evolving picture of the contribution of $s$-quarks to the proton's electric and magnetic form factors \cite{Ito04,Spayde04,G0typical, HAPPEx1,HAPPEx2,acha06}.

\subsection{Nucleon electromagnetic form factors}
\label{sec:ffintro}

In a Poincar\'e covariant treatment an on-shell $J^{P}= \frac{1}{2}^+$ nucleon with four-momentum $P$, mass $M$ and polarisation $s$ is described by a four-component spinor (column-vector) $u(P,s)$ that satisfies the Dirac equation:
\begin{equation}
(\slash \!\!\!\! P - M) u(P,s) = 0\,.
\end{equation}
In the nucleon's rest frame, $s^\mu = (0,\vec{s})$ with $\vec{s}\cdot\vec{s}=1$.  The adjoint spinor is $\bar u(P,s) = u(P,s)^\dagger \gamma^0$\, and we choose to normalise such that $\bar u(P,s^\prime) u(P,s) = 2 M \delta_{s^\prime s}$.  (NB.\, Our Minkowski space metric and Dirac matrix conventions are those of \cite{bd1,bd2}.)

The electromagnetic current for such a nucleon is 
\begin{eqnarray}
\label{Jnucleon}
J_\mu(P^\prime,s^\prime;P,s) & = & \bar u(P^\prime,s^\prime)\, \Lambda_\mu(q,P) \,u(P,s)\,, \\
& = &  \bar u(P^\prime,s^\prime)\,\left( \gamma_\mu F_1(q^2) +
\frac{1}{2M}\, i\sigma_{\mu\nu}\,q_\nu\,F_2(q^2)\right) u(P,s)\,,
\label{JnucleonB}
\end{eqnarray}
where $P,s$ ($P^\prime,s^\prime$) are the four-momentum and polarisation of the incoming (outgoing) nucleon and $q= P^\prime - P$ is the momentum transfer.  The quantity $\Lambda_\mu(q,P)$ in (\ref{Jnucleon}) is the fully-amputated nucleon-photon vertex, which in QCD is an eight-point Green function.  Poincar\'e covariance entails that the general expression for $\Lambda_\mu(q,P)$ involves twelve independent scalar functions but when projected via on-shell nucleon spinors, as in (\ref{Jnucleon}), all
tensor structures reduce to the two shown in (\ref{JnucleonB}) with the Poincar\'e-invariant scalar coefficient functions $F_1$ and $F_2$, which are termed, respectively, the Dirac and Pauli form factors.  
It is sometimes useful to work with the isoscalar and isovector combinations of these form factors
\begin{equation}
\label{isovectorF}
F_i^s=\frac{1}{2}\left(F_i^p + F_i^n\right),\  
F_i^v=\frac{1}{2}\left(F_i^p - F_i^n\right),\  (i=1,\,2).
\end{equation}

Two important combinations of the Dirac and Pauli form factors are the
so-called Sachs form factors \cite{ernst60}:
\begin{equation}
\label{GEpeq}
\ge(Q^2)  =  F_1(Q^2) - \frac{Q^2}{4 M^2} F_2(Q^2)\,,\; 
\gm(Q^2)  =  F_1(Q^2) + F_2(Q^2)
\end{equation}
($-q^2=Q^2>0$ defines spacelike), which express the nucleon's electric and magnetic form factors.  In the Breit frame and in the nonrelativistic limit, the three-dimensional Fourier transform of $\ge(Q^2)$ provides the electric-charge-density distribution within the nucleon, while that of $\gm(Q^2)$ gives the magnetic-current-density distribution \cite{sachs62}.  Naturally, $\gep(0)=1$, $\gen(0)=0$, which expresses the proton and neutron electric charges.  Moreover, $\gm(0)=(\ge(0)+\kappa)=:\mu$ defines the proton and neutron magnetic moments.  In this expression $\kappa=F_2(0)$ is the \emph{anomalous} magnetic moment, discovered for the proton by Stern: $\kappa_p=1.79$ in (\ref{stern}).  It is noteworthy, as we said, that for a point particle $F_2=0$, in which case $\ge=\gm$.

As reviewed, e.g., in \cite{gao03,HAPPEx}, it is possible to expose the contribution of individual quark flavours to these form factors by considering the coupling of the $Z^0$-boson to the nucleon.  This is expressed via the nucleon's neutral weak current
\begin{eqnarray}
\label{JZnucleon}
\rule{-13ex}{0ex}\lefteqn{J_\mu^{Z}(P^\prime,s^\prime;P,s) =  \bar u(P^\prime,s^\prime)\, \Lambda_\mu^Z(q,P) \,u(P,s)}\\
&  & \rule{-13ex}{0ex} =  \bar u(P^\prime,s^\prime)\,\left( \gamma_\mu F_1^Z(Q^2) + 
\frac{i\sigma_{\mu\nu}\,q_\nu}{2M}\, \,F_2^Z(Q^2) + \gamma_5\gamma_\mu G_A(Q^2) + \gamma_5 \frac{q_\mu}{M} G_P(Q^2)\right) u(P,s)\,, \label{JZnucleonB}
\end{eqnarray}
where the new scalar functions appearing in (\ref{JZnucleonB}) are the axial-vector, $G_A$, and pseudoscalar, $G_P$, nucleon form factors, associated with the axial-vector part of the $Z^0$-nucleon coupling, which herein we cannot discuss further.

The contribution from each quark flavour to a given form factor is defined by writing that form factor as a sum in which each of the terms is multiplied by the relevant electric or weak quark charge; viz., for the proton's electric and magnetic form factors,
\begin{eqnarray}
G_{E,M}^p(Q^2) & = & \frac{2}{3} \, G_{E,M}^{pu}(Q^2) - \frac{1}{3} \left[ G_{E,M}^{pd}(Q^2) + G_{E,M}^{ps}(Q^2)\right], \\
\nonumber
G_{E,M}^{Zp}(Q^2) & = & (1 - \sfrac{8}{3}\sin^2\theta_W) \, G_{E,M}^{pu}(Q^2)\\
&&  - (1 - \sfrac{4}{3}\sin^2\theta_W) \left[ G_{E,M}^{pd}(Q^2) + G_{E,M}^{ps}(Q^2)\right],
\end{eqnarray}
where $\theta_W$ is the weak-mixing angle: $\sin^2\theta_W(M_Z) \approx 0.23$.
Here the form factors are the same in each line because, once the charges are factorised, the matrix elements are constructed from the same quark-level vector current whether the probe is electromagnetic or weak-vector.  The contribution from quarks heavier than strange is supposed to be small.

If one assumes charge symmetry then the $d$-quark contribution to the neutron's form factors is the same as the $u$-quark contribution to the proton's; i.e., $G_{E,M}^{nd}= G_{E,M}^{pu}$, $G_{E,M}^{nu}= G_{E,M}^{pd}$, and the $s$-quark contribution is the same in both the proton and neutron; viz., $G_{E,M}^{ps}=G_{E,M}^{ns}=G_{E,M}^s$.  Hence
\begin{equation}
\label{lastGEs}
G_{E,M}^{Zp}(Q^2) = (1 - 4\sin^2\theta_W) G_{E,M}^p(Q^2) - G_{E,M}^n(Q^2) - G_{E,M}^s(Q^2),
\end{equation}
so that the $s$-quark contributions to the electric and magnetic form factors
are determined once one has $G^p$, $G^n$ and $G^Z$.  The latter is accessible
via parity violating electron scattering from the proton, which is covered at
length in \cite{HAPPEx}.

Form factors are truly an important gauge of a hadron's structure.  They are a measurable and physical manifestation of the nature of the hadron's constituents and the dynamics that binds them together.  In analogy with optics, the scattering of a probe with three-momentum $\vec{Q}$ should resolve the structure of the hadron on a length-scale $d\sim 1/|\vec{Q}|$.  Hence, so long as $|\vec{Q}|\ll M$; namely, recoil effects are small, then one has a straightforward interpretation of the data in terms of a static charge- and current-distribution \cite{kelly02,Ralston:2002ep,sick03,blunden05b,sick05} and, e.g., the proton's electric and magnetic radii are determined via
\begin{equation}
\label{genrad}
\langle r_{E,M}^2\rangle= \frac{- \, 6\,}{G_{E,M}(0)}\left[\frac{d}{dQ^2} \, G_{E,M}(Q^2) \right]_{Q^2=0}.
\end{equation}
[NB.\ Should $G(0)=0$, this normalising factor is omitted; e.g., (\ref{nradius}) and (\ref{rps}).]  Even though these aspects of the interpretation break-down for $Q^2\ngsim M^2$, the form factors remain as a visible probe and exacting test of our understanding of QCD's dynamics.



\subsection{Electron scattering formalism}
%
%
Electron--nucleon scattering is typically treated in the single-photon-exchange
approximation.  As intrinsic properties of the target, the nucleon form factors are independent of this approximation but their inference from experiment is not.  In the single-photon-exchange approximation the differential cross section for elastic electron-nucleon scattering can be expressed by the Rosenbluth \cite{rosenbluth50} formula
\begin{equation}
\frac{d\sigma}{d\Omega} = \sigma_M \left( F_1^2(Q^2) + \tau F_2^2(Q^2) +
2\tau [F_1(Q^2) + F_2(Q^2)]^2 \tan^2(\theta_e/2)\right)\ ,
\end{equation}
where $\tau=Q^2/(4M^2)$, $\sigma_M =(\frac{\alpha_{QED}\cos(\theta_e/2)}{2E_e\sin^2(\theta_e/2)})\frac{E_e^\prime}{E_e}$ is the Mott cross-section for scattering from a point-like particle, $\alpha_{QED}$ is the fine-structure constant, $E_e$ is the initial electron energy, and $E_e^\prime$ and $\theta_e$ are the scattered electron energy and angle, respectively.  Reference \cite{hand60} re-expressed the Rosenbluth formula in terms of $\ge$ and $\gm$; viz., 
\begin{equation}
\frac{d\sigma}{d\Omega} = \frac{\sigma_M}{\varepsilon (1+\tau)}
\left[ \tau \, \gm^2(Q^2) + \varepsilon \, \ge^2(Q^2) \right] ,
\label{eq:rosenbluth}
\end{equation}
where $\varepsilon=1/[1 + 2(1+\tau)\tan^2(\theta_e/2)]$ is the virtual photon
polarisation parameter.  


With polarised electron beams, one also can measure the cross section asymmetry from a polarised target, or the polarisation transfer to an unpolarised nucleon.  For a polarised nucleon target, the beam-target asymmetry is \cite{dombey69,akhiezer74}:
\begin{equation}
\rule{-4em}{0ex} A_{\vec{e}\vec{N}} = \frac{\kappa_1 \cos{\theta^*} \gm^2 + \kappa_2 \sin{\theta^*} \cos{\phi^*} \ge \gm} {\ge^2 + \kappa_3 \gm^2}
= \frac{\kappa_1 \cos{\theta^*} + \kappa_2 \sin{\theta^*} \cos{\phi^*} R}{R^2 + \kappa_3} ,
\label{eq:asymmetry}
\end{equation}
where $\kappa_{1,2,3}$ are kinematic factors, $\theta^*$, $\phi^*$ are the polar and azimuthal angles between the scattering plane and nucleon spin direction, and $R=\gegm$.  Note that in this case one can vary $\theta^*$ to isolate the transverse component ($\theta^* = 90\deg$), which is very sensitive to $R$, or the longitudinal component ($\theta^*=0$), which has little sensitivity to the form factors unless $R^2\ngsim \kappa_3$.  One can also measure the transverse and longitudinal polarisation transfer components which have similar sensitivity to $R$.

\subsection{Layout}
We have divided this article into five sections.  This closes the Introduction.  Section\;\ref{experimentJA} presents an experimental perspective, with an overview of the current status of nucleon form factor measurements.  Section\;\ref{theoryCDR} provides a snapshot of the impact that these measurements are having on the phenomenology and theory of hadron physics and QCD.  The challenges that a description of nucleon form factors poses for numerical simulations of lattice-regularised QCD and its successes are summarised in Section\;\ref{sec:lattice}.  Section\;\ref{Epilepsy} is an Epilogue. 

\section{Experimental status}\label{experimentJA}

Until very recently, most extractions of the nucleon elastic form factors came from unpolarised, inclusive electron scattering measurements.  While this provided a significant body of data on $\gmp$, measurements of $\gep$ were limited at high $Q^2$, and measurements of the neutron form factors were much less precise.

In the last decade, high intensity beams, large and efficient neutron detectors, and high polarisation beams and targets have led to a dramatic improvement in our knowledge of the electromagnetic form factors.  These new experiments, coupled with a better understanding of higher order electromagnetic corrections (specifically, two-photon exchange effects), have led to a dramatic increase in the kinematic coverage, precision and completeness of the form factor data.  In particular, there will soon be measurements of all four elastic electromagnetic form factors up to $Q^2
\approx 3.5$~GeV$^2$, allowing for comparisons with nucleon structure models at both low $Q^2$, where the pion cloud is believed to play an important role, and large $Q^2$, where one is sensitive to the nucleon's quark core.  In addition, comparisons of the proton and neutron form factors enable one to make model-independent statements about flavour-dependent aspects of the nucleon.  Since contemporary simulations of lattice-regularised QCD cannot handle disconnected diagrams, this also provides access to the isovector form factors for which lattice calculations are currently feasible. 

\subsection{Unpolarised elastic and quasi-elastic scattering}
Prior to the advent of polarisation transfer measurements, our knowledge of the proton form factors came almost entirely from inclusive elastic $e$--$p$ scattering, with the form factors extracted using the Rosenbluth separation technique that relies on the simple dependence of the reduced cross-section, $\sigma_R$, on the virtual photon polarisation parameter, $\varepsilon$ [see (\ref{eq:rosenbluth})]:
\begin{equation}
\sigma_R \equiv \frac{d\sigma}{d\Omega} \frac{\varepsilon(1+\tau)}{\sigma_{\rm Mott}}
= \tau \, \gm^2(Q^2) + \varepsilon \, \ge^2(Q^2),
\label{eq:sigr}
\end{equation}
allowing one to extract $\gm$ from $\sigma_{ep}$ at $\varepsilon=0$ and $\ge$ from the $\varepsilon$ dependence.  These extractions have several important limitations; viz., reduced sensitivity to $\ge$ at large $Q^2$ and $\gm$ at small $Q^2$ (except for $\theta \rightarrow 180\deg$), a large anti-correlation between the errors in the extracted values of $\ge$ and $\gm$, and a significant correlation between the values of $\ge$ (or $\gm$) extracted at different $Q^2$ from a single experiment owing to uncertainties in the $\varepsilon$-dependent corrections applied to the data.


\begin{figure}[t]
\begin{center}
\epsfig{height=3.0in,width=4.0in,file=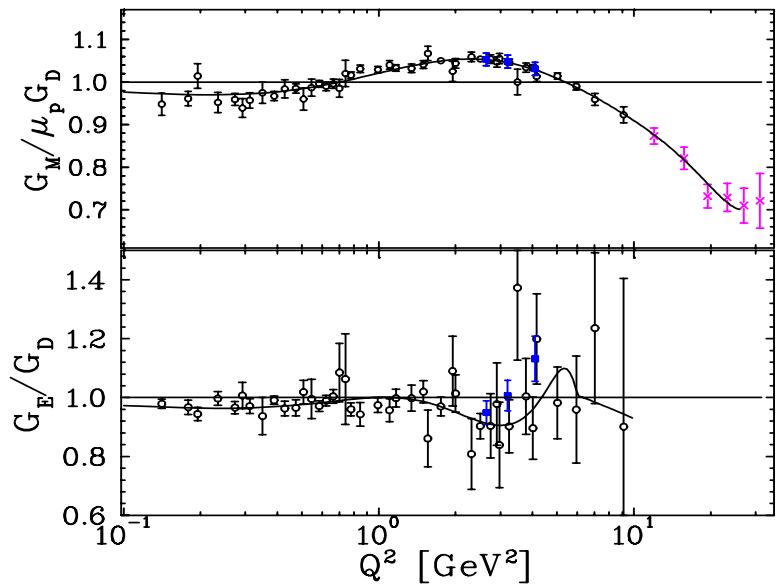}
\end{center}
\caption{Ratio of $\gep$ and $\gmp$ to the dipole form: $1 / (1 + Q^2/m_D^2)^2$, $m_D^2=0.71\,$GeV$^2$.  The hollow circles and solid curve show the global analysis of the cross-section data from \cite{arrington04a}.  The solid squares are the results from \cite{qattan05}, and the crosses are the high-$Q^2$ measurements from \cite{sill93}, which were not included in~\cite{arrington04a}.
\label{fig:ros_proton}}
\end{figure}

Despite these limitations, data from such measurements were sufficient to provide high precision extractions of $\gmp$ for $Q^2\in [0.1,30]\,{\rm GeV}^2$, and $\gep$ from 0.01--2~GeV$^2$ (see Figure~\ref{fig:ros_proton}).  The uncertainties on $\gep$ grow rapidly with $Q^2$ owing to the reduced contribution of $\gep$ to $\sigma_R$.  Both form factors are reasonably well approximated by a simple dipole form, $\gep = \gmp/\mu_p = 1 / (1 + Q^2/0.71)^2$, with $Q^2$ in GeV$^2$, although the $\gmp$ data are precise enough to show clear deviations from this dipole fit.  The $\gep$ data are systematically below the dipole fit on a sizeable domain near 0.1~GeV$^2$, but the data at large $Q^2$ do not exhibit deviations of the magnitude seen in $\gmp$. 

Similar measurements of the neutron form factors are extremely limited.  Extracting $\gen$ from unpolarised scattering is nearly impossible because the small value of $\gen$ provides at most 5--6\% of the total $e$--$n$ cross-section.  While this makes extraction on $\gmn$ easier, the absence of a free neutron target is still a significant limitation.  Most experiments extracted $\gmn$ using inclusive quasi-elastic electron-nucleon scattering from deuterium \cite{bartel72,hanson73, bartel73, rock82, esaulov87, arnold88,lung93}.  This requires subtracting the contribution coming from quasi-elastic $e$--$p$ scattering, yielding significant uncertainties that are correlated with the knowledge of the proton form factors.  Such extractions also require corrections for the nuclear effects in deuterium.  One experiment extracted $\gmn$ from a coincidence $D(e,e^\prime,n)$ measurement \cite{markowitz93}.  By tagging the struck neutron, one removes the contamination from $e$--$p$ scattering, but becomes sensitive to knowledge of the neutron detection efficiency and the much larger nuclear effects involved when detecting the neutron as well as the electron.  These experiments provided measurements of $\gmn$ with $\sim$3\% uncertainties on $Q^2 \nlsim 0.5\,$GeV$^2$, $\gtorder 5$\% uncertainty for 0.5--5\,GeV$^2$, and 10-15\% uncertainty on  $5\nlsim Q^2\nlsim$10 GeV$^2$.



For $\gen$, only upper limits could be set by such measurements.  At very low $Q^2$, neutron--electron scattering measurements have been used to extract the neutron charge radius~\cite{koester95}, which is defined via
\begin{equation}
\label{nradius}
\langle r_n^2 \rangle = \left. -6 \frac{d}{dQ^2} G_E^n(Q^2)\right|_{Q^2=0},
\end{equation}
because $G_E^n(0)=0$, cf.\ (\ref{genrad}).  Prior to recent polarisation measurements, the best extraction of $\gen$ at finite $Q^2$ came from analyses of elastic $e$--$D$ scattering, where a model-dependent extraction of $\gen$ can be obtained. This extraction is extremely sensitive to the $N$--$N$ potential used in calculating the deuterium wave function.  Thus, while these measurements give some indication of the $Q^2$ dependence of $\gen$, the values extracted in different analyses vary by up to a factor of two~\cite{galster71}.

\subsection{Polarised elastic scattering and ratio measurements}
The difficulties in extracting form factors from unpolarised scattering, owing to lack of a free neutron target and reduced sensitivity to $\ge$ at high $Q^2$, made it necessary to employ new techniques.  These techniques had been known for some time, but required improved beams and/or detectors, which have only recently become routinely available.

Many techniques had been proposed to make improved measurements of the nucleon form factors.  For $\gmn$, measurements of $D(e,e^\prime,n)$ could be used to exclude proton scattering contributions, while a comparison of $D(e,e^\prime,n)$ with $D(e,e^\prime,p)$ significantly reduces the importance of nuclear corrections \cite{durand59}.  Such measurements require a combination of high luminosity and large or efficient neutron detectors.  These have become available in recent years and several such measurements have been performed \cite{bruins95,anklin94, anklin98, kubon02,lachnietphd}.  These experiments yield significantly more precise measurements of $\gmn$, reducing the uncertainties by roughly a factor of two up to $Q^2=1$~GeV$^2$.  In the ratio $D(e,e^\prime,n)/D(e,e^\prime,p)$, nuclear corrections largely cancel and subtraction of a large proton contribution is not required. 


\begin{figure}[t]
\begin{center}
\epsfig{angle=90,width=3.0in,height=2.2in,file=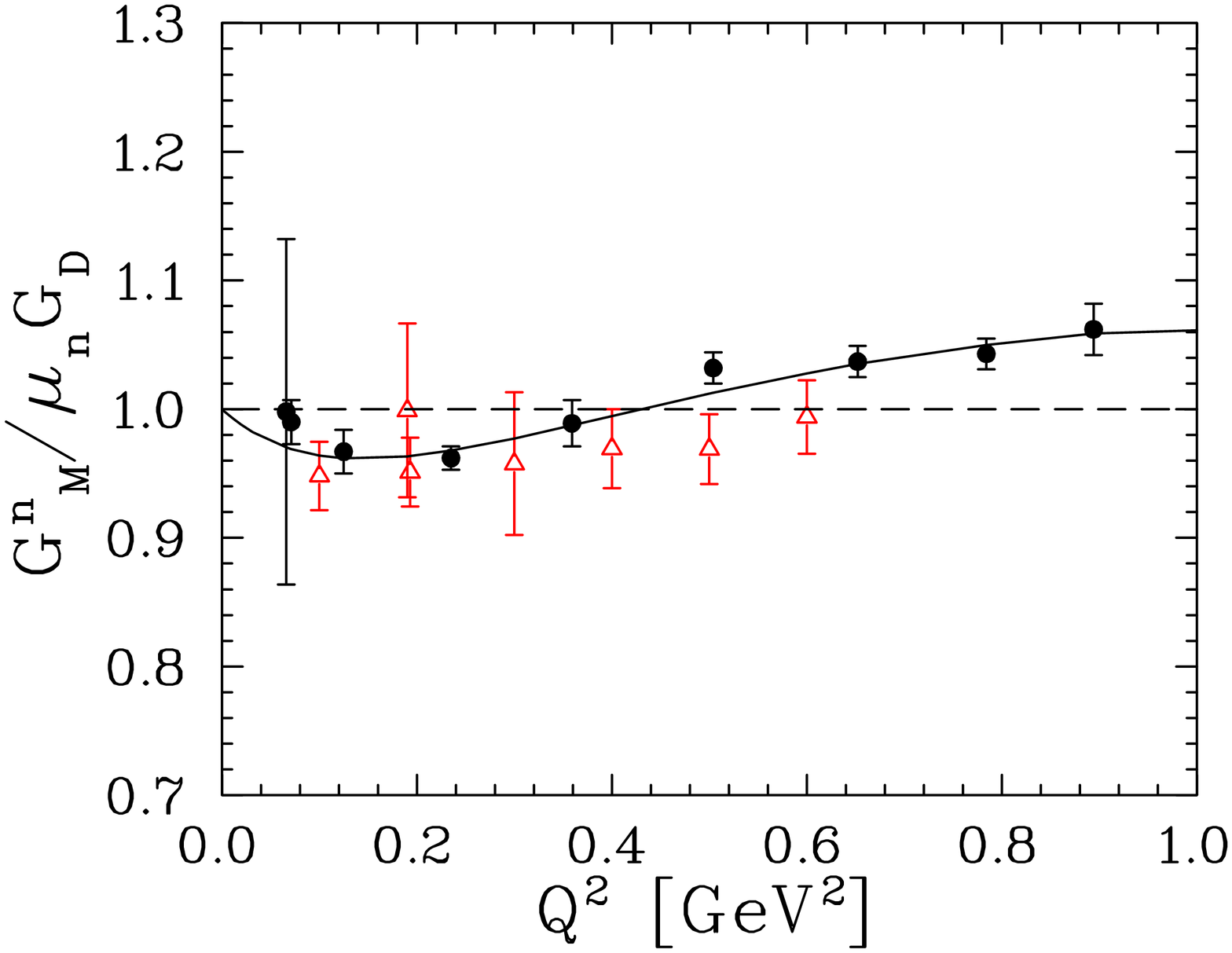}
\epsfig{angle=90,width=3.0in,height=2.2in,file=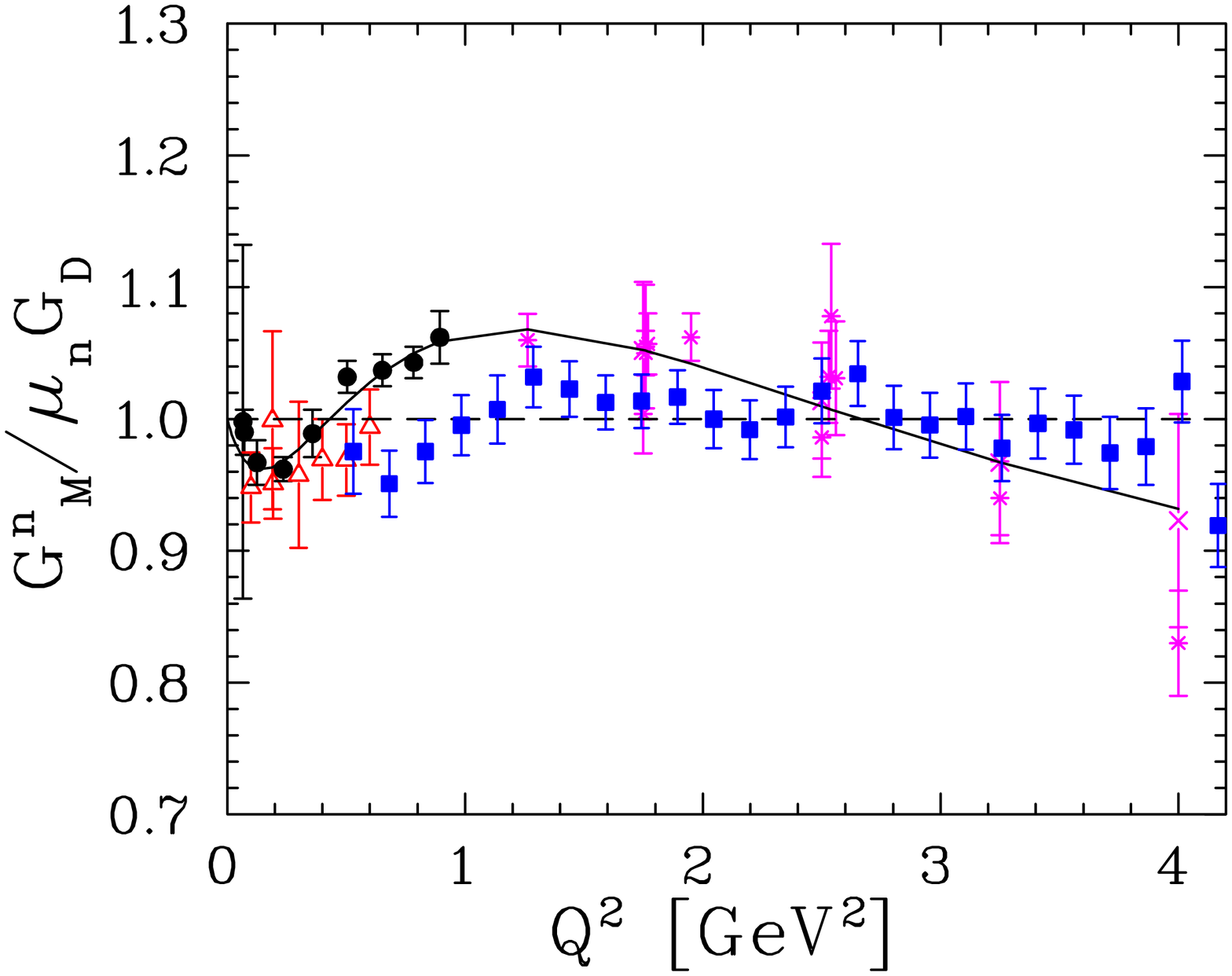}
\end{center}
\caption{Values of $\gmn$ taken from ratio measurements on deuterium and polarised $^3$He measurements.  The circles are extractions from the ratio of $e$--$n$ to $e$--$p$~\ quasielastic scattering and the open triangles are from measurements on polarised $^3$He.  The right panel includes high $Q^2$ data: the solid squares are the CLAS preliminary results \cite{lachnietphd}, and the crosses \cite{lung93} and asterisks \cite{rinat04} indicate points taken from quasielastic $e$--$n$ scattering on light nuclei.  The solid line is a fit from \cite{kelly02}.
\label{fig:gmn}}
\end{figure}

Polarisation measurements have also provided improved extractions of $\gmn$.  For polarised $e$--$n$ scattering, both the parallel and perpendicular asymmetry are sensitive only to the \textit{ratio} $\gegm$ in scattering from a free nucleon.  However, the technique can be used to extract $\gmn$ in quasi-elastic scattering from polarised $^3$He \cite{blankleider83}.  In a simplified picture, neglecting nuclear effects and assuming unpolarised protons, the scattering is a combination of scattering from two unpolarised protons and one polarised neutron.  The parallel beam-target asymmetry, $\theta^\ast=0$ in (\ref{eq:asymmetry}), from the neutron is very well known because it depends only on the kinematics and a small correction from $(\gegmn)^2$.  Since the neutron asymmetry is well known, the experiment is essentially a measurement of the dilution of the asymmetry coming from the contribution of the unpolarised protons. Thus, it is essentially a measurement of the ratio of $\sigma_{en}$ to $\sigma_{ep}$.  As with the direct ratio measurements on deuterium, this yields minimal sensitivity to the proton form factors. The analysis of these experiments~\cite{gao94, anderson06} takes into account nuclear effects, the polarisation of the protons, and all other effects neglected in the simple picture.  Figure~\ref{fig:gmn} shows the extracted values of $\gmn$ from ratio measurements on deuterium and the quasielastic asymmetry in polarised $^3$He.

\begin{figure}[t]
\vspace{-0.4cm}
\begin{center}
\epsfig{angle=0,width=3.0in,height=2.2in,file=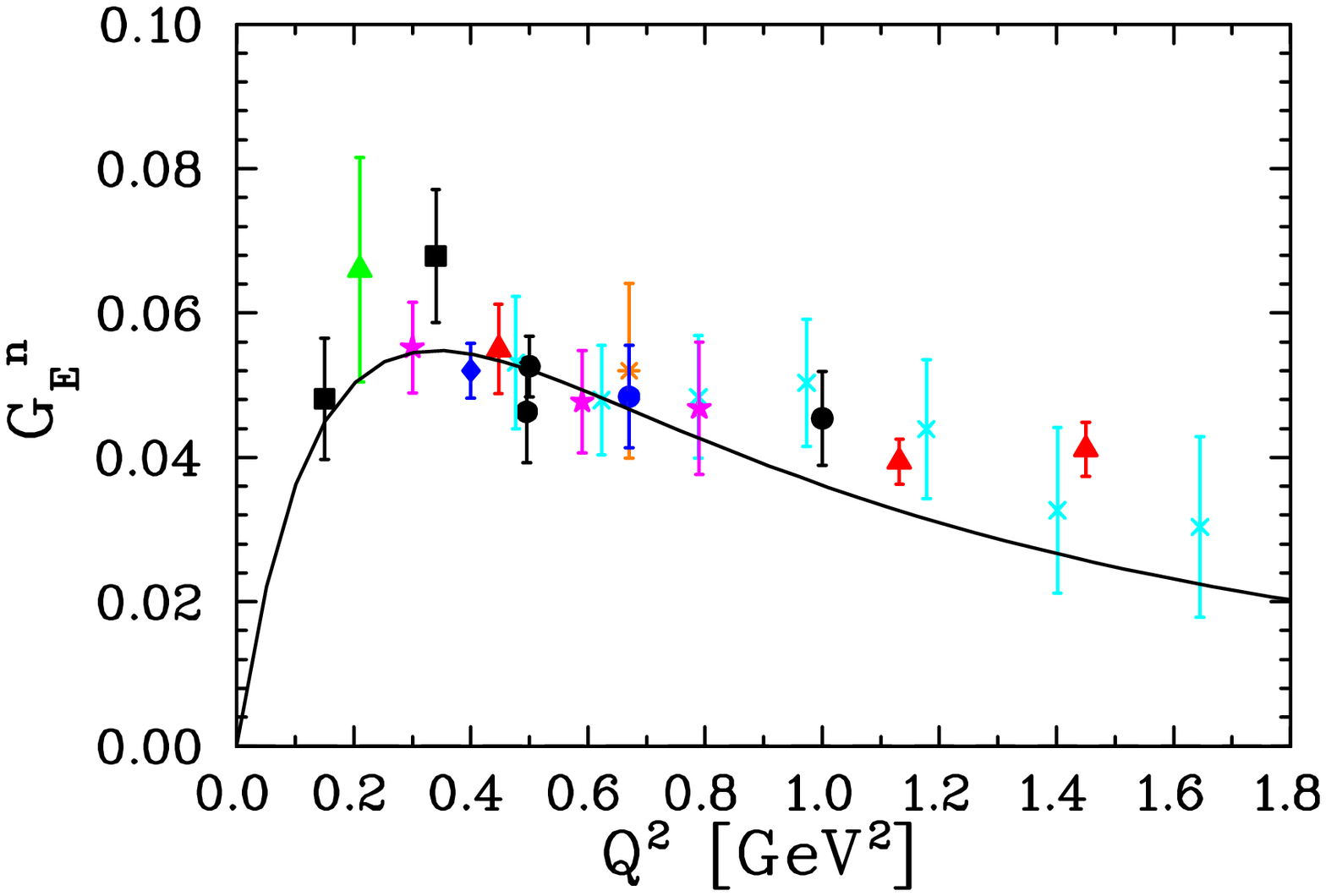}
\epsfig{angle=0,width=3.0in,height=2.2in,file=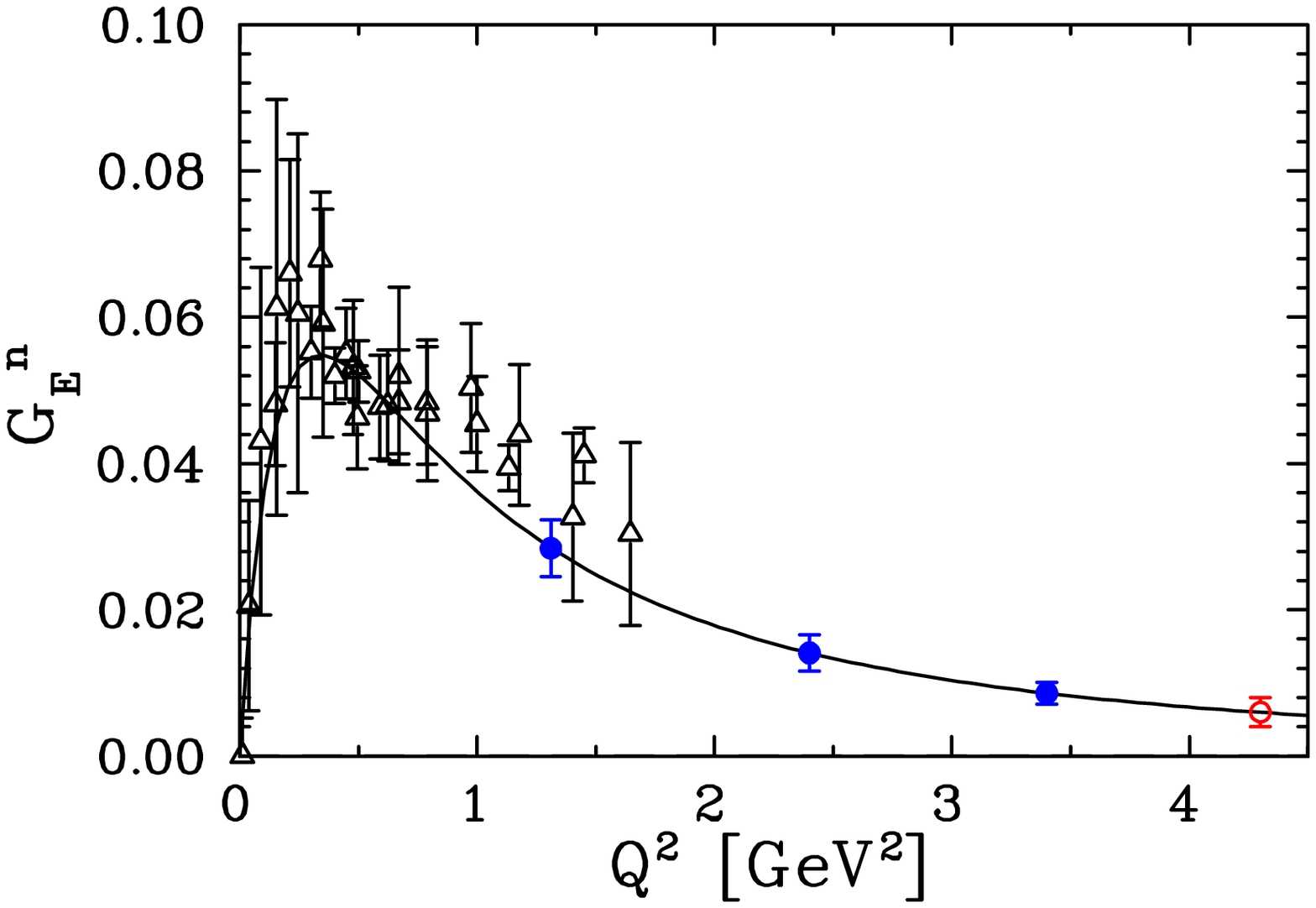}
\end{center}
\caption{Values of $\gen$ taken from \cite{passchier99}--\cite{schiavilla01}, compared to the Galster parametrisation \cite{galster71}.  The right panel includes projected uncertainties for future high-$Q^2$ measurements: a recently completed JLab Hall A measurement (solid circles), and an approved Hall C measurement (hollow circle).
\label{fig:gen}}
\end{figure}

Owing to their sensitivity to $\gegm$, measurements of spin-dependent $e$--$n$ scattering, via polarisation transfer or beam-target asymmetry measurements, have  been used more extensively to extract $\gen$.  The perpendicular asymmetry, $\theta^\ast=90^\circ$ and $\phi^\ast=0$ in (\ref{eq:asymmetry}), is sensitive to the ratio $\gegm$ \cite{dombey69,akhiezer74}.  It provides a better way to measure $\gen$ than the Rosenbluth separation technique, where the relative contribution of $\gen$ is suppressed as $(\gegmn)^2$.  Measurements of the asymmetry have been made using quasielastic scattering from polarised deuterium \cite{passchier99, warren04} and $^3$He~\cite{meyerhoff94,rohe99, golak01,bermuth03} targets, as well as measurements of recoil polarisation from an unpolarised deuterium target \cite{eden94,herberg99, madey03,glazier05}.  Figure~\ref{fig:gen} shows the results of the $\gen$ measurements described above.  Also shown (as crosses) are the $Q^2 > 0.4$~GeV$^2$ results from a more recent extraction of $\gen$ via deuterium form factor measurements \cite{schiavilla01}.


Finally, polarisation measurements have also been used to improve high-$Q^2$ measurements of $\gep$, where the cross-section is dominated by the contribution from $\gmp$.  Two measurements used the cross-section asymmetry on polarised proton targets \cite{jones06, crawford06}, but most measurements, including all the high-$Q^2$ extractions, have used recoil polarisation \cite{punjabi05, gayou02, milbrath99,gayou01, pospischil01,maclachlan06}.  Along with taking advantage of the increased sensitivity to $\gep$ in the transverse polarisation transfer (or perpendicular asymmetry), some experiments have made simultaneous extractions of the longitudinal and transverse polarisation transfer.  By taking the ratio of these components, one maintains sensitivity to $\gegmp$ while cancelling the uncertainties associated with the beam polarisation and the analysing power of the recoil polarimeter.  Combining measurements with positive and negative helicity electrons, false asymmetries in the recoil polarimeter also cancel, yielding an extraction of $\gegmp$ with extremely small systematic uncertainties \cite{punjabi05}.

\begin{figure}[t]
\begin{center}
\epsfig{angle=0,width=3.8in,height=2.6in,file=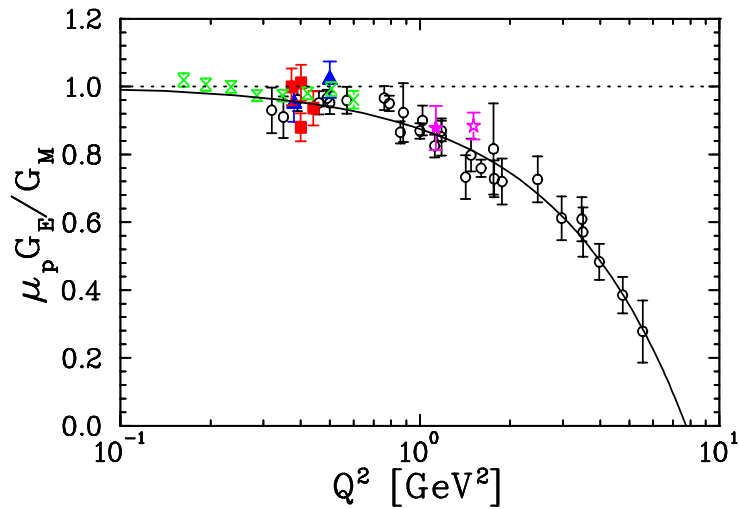}
\end{center}
\caption{Polarisation measurements of $\mugegmp$, showing the combined
statistical and systematic uncertainties.  The recoil polarisation data are
shown as solid triangles \cite{milbrath99}, squares \cite{pospischil01},
stars \cite{maclachlan06}, and hollow circles \cite{punjabi05,gayou02,gayou01}.  The polarised target measurements are shown as hollow
stars~\cite{jones06}, and crosses~\cite{crawford06}. The solid line is a
linear fit to the JLab data, $\mugegmp=1-0.13(Q^2-0.04)$ \cite{gayou02}.
\label{fig:gep}}
\end{figure}

Results from the polarisation transfer measurements of $\gegmp$ are shown in Figure~\ref{fig:gep}.  These measurements indicate a significant decrease in the ratio at large $Q^2$ and suggest that $\gep$ may possess a zero at somewhat higher $Q^2$.  This differs markedly from conclusions based on Rosenbluth separation extractions, which indicated that $\gegmp$ was roughly consistent with unity up to $Q^2\sim 6\,$GeV$^2$.  Notably, the precise data from BLAST \cite{crawford06} have somewhat smaller uncertainties than a global analysis of low-$Q^2$ cross-section data \cite{sick03}.  Hence, while in a combined analysis at low-$Q^2$ of the cross-section and polarisation data the extraction of $\gep$ and $\gmp$ will remain limited by normalisation uncertainties in the cross-section measurements, the addition of the BLAST data will significantly reduce the anti-correlation between the extracted values of $\gep$ and $\gmp$.

When the discrepancy in Figure~\ref{fig:gep} was first observed, it was noted that there is significant scatter in the values of $\gep$ extracted from different Rosenbluth measurements, especially at $Q^2$ values above 1--2~GeV$^2$, which is not seen in the combined analysis depicted in Figure~\ref{fig:ros_proton}.  Owing to this scatter it was often assumed that the discrepancy between Rosenbluth and polarisation measurements arose from systematic uncertainties in the Rosenbluth extractions.  However, the scatter is largely explained by the fact that many experiments extracted the form factors by combining new cross-section data with results from previous measurements.  Differences in the relative normalisation of low-$\varepsilon$ and high-$\varepsilon$ data sets leads to significant errors in the extracted values of $\gep$ that affect the entire $Q^2$ range of the analysis.  The normalisation uncertainties were often ignored entirely, or else estimated but treated as uncorrelated.  It was demonstrated \cite{arrington03a} that the various data sets were consistent when performing Rosenbluth separations using only data from individual experiments, or when properly taking the normalisation uncertainties into account.  More recently, the discrepancy was demonstrated much more clearly by a new, high-precision Rosenbluth extraction of $\gegmp$ \cite{qattan05}.  By detecting the struck proton, rather than the scattered electron, this experiment had significantly smaller $\varepsilon$-dependent corrections, allowing for smaller systematic uncertainties.

Thus, the inconsistency implies either an error in the polarisation measurements, a common systematic error in the analysis of the cross-sections, or a fundamental flaw in one of the two techniques.  Even if one assumes that the problem lies with the Rosenbluth data, owing to the greater sensitivity of the polarisation measurements, it is important to fully understand the discrepancy.  This because the cross-section data are needed to separate $\ge$ and $\gm$, and both the elastic cross-section and form factor measurements are important ingredients in other analyses.  The source of the difference will determine its impact on other measurements \cite{arrington04a}.  Currently, the discrepancy is believed to be explained by two-photon exchange corrections, which are discussed in Section~\ref{sec:twophoton}.

\subsection{Future form factor measurements}
\label{futureexperiment}
In the next few years final results should become available from the CLAS $\gmn$ measurements at JLab \cite{brooks05}, the $\gen$ experiments in Hall A \cite{e02013} and the BLAST experiment at MIT-Bates.  Additional data on $\gmn$ were taken using a deuterium target, with a low momentum spectator proton tagged to give a nearly-free neutron target \cite{e03012}.  An extension of the polarisation transfer measurement of $\gegmp$ to $Q^2=8.5$~GeV$^2$ will be performed in 2007 \cite{e04108}, along with a series of high-precision Rosenbluth extractions of $\gep$ and $\gmp$ \cite{e05017}. In the longer term, the energy upgrade at Jefferson Lab will allow high precision measurements of $\gegmp$ and $\gmn$ to $Q^2 \approx 14$~GeV$^2$, as well as an extension of $\gen$ measurements to 8~GeV$^2$.

\subsection{Flavour decomposition, parity violating scattering}
\label{expflavour}
The combination of proton and neutron form factors can be used to study the flavour dependence of the charge and magnetisation distributions.  Combining proton and neutron measurements provides sensitivity to the difference between up and down quark distributions because of the difference in relative weighting between the proton and neutron.  To capitalise on this, one needs data for $\gep$ and $\gen$ (or $\gmp$ and $\gmn$) that cover the same $Q^2$ range and have comparable precision.  With the recent and upcoming measurements of neutron form factors, one will have measurements of all four form factors at $Q^2$ values up to $\approx$5~GeV$^2$.  The neutron measurements have larger relative uncertainties, but the absolute uncertainties for the proton and neutron measurements are comparable.  This provides an optimal case for comparison of neutron and proton measurements, which can be used to study the flavour dependence.

Adding measurements of parity violating elastic scattering allows one to fully separate the $u$, $d$, and $s$ quark contributions, (\ref{JZnucleon})--(\ref{lastGEs}).  However, in addition to requiring this new data, a careful treatment of the correlations between the form factor measurements is required as well as calculations of the two-photon-exchange (TPE) corrections, Section~\ref{sec:twophoton}.  While the size of TPE corrections in the parity violating asymmetry is small \cite{afanasev05b}, the impact of TPE in extracting the electromagnetic form factors is large enough that is must be accounted for in a combined analysis of parity-conserving and parity-violating elastic scattering~\cite{arrington06b}, especially for $Q^2\ngsim 1\,$GeV$^2$.

\section{Theoretical understanding}
\label{theoryCDR}
\subsection{Background}
\label{sub:back}
The experimental results reviewed in Section~\ref{experimentJA} are the objective facts for which strong interaction theory must provide an understanding.  To this we now turn.

At large spacelike $Q^2$, perturbative QCD amplitudes factorise.  It follows that helicity is conserved at leading twist and hence for $Q^2=-q^2 > \zeta_{\rm pQCD}^2 \gg \Lambda_{\rm QCD}^2$
\begin{equation}
\label{F2F1scale}
\frac{Q^2 F_2(Q^2)}{F_1(Q^2)} \approx\, {\rm constant}\,;
\end{equation}
viz., the Pauli form factor is power-law suppressed with respect to the Dirac because it is an helicity-flip amplitude~\cite{bl81}.  Indeed, dimensional counting rules for QCD's hard amplitudes give~\cite{Brodsky:1973kr,Matveev:1973ra, Brodsky:1974vy} $F_1(Q^2) \sim 1/Q^4$, $F_2(Q^2) \sim 1/Q^6$, and hence
\begin{equation}
\label{GEGMscale}
\frac{\ge(Q^2)}{\gm(Q^2)} \stackrel{Q^2 >\, \zeta_{\rm pQCD}^2}{\sim} \,{\rm constant}.
\end{equation}
Perturbative QCD and dimensional counting rules cannot predict the value of
$\zeta_{\rm pQCD}$; namely, the scale at which these results should become
evident in experiment.  That requires a nonperturbative method.  However, the
experiments which form the basis of Figure\,\ref{fig:gep} suggest strongly that
$\zeta_{\rm pQCD}^2 \gg 6\,$GeV$^2$.  Moreover, a linear fit to the polarisation transfer data yields a zero in $\gep(Q^2)$ at $Q^2\approx 7.8\,$GeV$^2$, while $\gmp(Q^2)$ remains positive definite.  The possibility of a zero in $\gep(Q^2)$ was largely overlooked before the polarisation transfer data became available, although thereafter many models were found to exhibit such behaviour.  The confirmation of a zero is sought in a forthcoming JLab experiment \cite{e04108}, which will obtain polarisation transfer data out to $Q^2=8.5$~GeV$^2$.  

While a zero in $\gep(Q^2)$ was not generally anticipated, a ratio $R_p(Q^2)= \mu_p \gep(Q^2)/\gmp(Q^2)$ that falls with increasing $Q^2$ could early have been inferred from vector meson dominance fits to existing data \cite{Iachello:1972nu}.  In a later dispersion-relation fit to electron-nucleon scattering cross-sections \cite{Hohler:1976ax} the ratio was explicitly calculated with an aim to testing (\ref{GEGMscale}) and found to fall with increasing $Q^2$ on a domain $0.5\nlsim Q^2({\rm GeV}^2)\nlsim 3.0$.  This behaviour persists in more recent analyses of this type \cite{Lomon:2001ga,Bijker:2004yu,Belushkin:2006qa}.  

\subsection{Mean-field and potential models}
At its simplest, the nucleon is a nonperturbative three-body bound-state problem, an exact solution of which is difficult to obtain even if the interactions are known.  Hitherto, therefore, phenomenological mean-field models have been widely employed to describe nucleon structure; e.g., soliton models~\cite{Birse:1991cx,Alkofer:1995mv,Schechter:1999hg} and constituent-quark models~\cite{Thomas:1982kv,Miller:1984em,Capstick:2000qj}.  Such models are most naturally applied to processes involving small momentum transfer ($Q^2< M^2$, $M$ is the nucleon mass) but, as commonly formulated, their applicability may be extended to processes involving moderately larger momentum transfer by working in the Breit frame \cite{Miller:1997jr}.  An alternative is to define an equivalent, Galilean invariant Hamiltonian and reinterpret that as the Poincar\'e invariant mass operator for a quantum mechanical theory~\cite{Coester:1992cg,Coester:2001eb}.  This is likely a truer extension of a model to large-$Q^2$.

In the context of soliton models, a zero in $\gep(Q^2)$ at $Q^2\sim 10$~GeV$^2$ was evident in a study of nucleon electromagnetic form factors based on a topological model that incorporated coupling to vector mesons and relativistic recoil corrections~\cite{Holzwarth:1996xq}.  However, owing to the uncertainties noted above in treating relativistic recoil corrections, it was argued therein that the predictive power of the model was poor at large-$Q^2$.  

Appropriate to the study of electromagnetic form factors at $Q^2> M^2$, \cite{Miller:2002qb} employed a light-front constituent-quark model with a wave-function parametrisation introduced in~\cite{Chung:1991st,Schlumpf:1992vq}.  In an earlier application, this model was found to produce a proton electric form factor with a zero at $Q^2\approx 5.7\,$GeV$^2$ \cite{Frank:1995pv}.  Although the predicted position of the zero is contradicted by existing data, it is a feature of this type of model that a zero in $\gep(Q^2)$ is easily obtained.  

The data in Figure\,\ref{fig:gep} can be re-expressed as a measure of $\kappa_p F_2^p(Q^2)/F_1^p(Q^2)$.  This ratio is approximately constant on \mbox{$2 \nlsim  Q^2 ({\rm GeV}^2) \nlsim 6$}.  Such behaviour has been argued to indicate the presence of substantial quark orbital angular momentum in the proton \cite{Ralston:2002ep,Miller:2002qb,Afanasev:1999at,Ralston:2003mt,Bloch:2003vn}.  It is a feature of \cite{Miller:2002qb} that $\sqrt{Q^2} F_2^p(Q^2)/F_1^p(Q^2)\approx\,$constant for $2 \nlsim Q^2 ({\rm GeV}^2) \nlsim 20$.  However, such behaviour is not uniformly found.  Moreover, it has been argued \cite{Brodsky:2003pw} that the dependence of form factors on $\sqrt{Q^2}$ instead of $Q^2$ is an artefact of the construction employed in~\cite{Chung:1991st,Schlumpf:1992vq}.  The lack of analyticity in $Q^2$ can be related to a violation of crossing symmetry.  Crossing symmetry is a necessary property of quantum field theory.  Its violation in the model of refs.~\cite{Miller:2002qb,Chung:1991st,Schlumpf:1992vq,Frank:1995pv,Miller:2002ig} indicates that while the model is relativistic and may provide an efficacious tool for computing low-$Q^2$ properties of hadrons, it is not applicable at large momentum transfer because it is inconsistent with essential features of quantum field theory.

A comparative analysis of relativistic constituent-quark model calculations of baryon form factors is provided in \cite{Julia-Diaz:2003gq}.  Model results depend in general on the representation of the baryon mass operator, which is expressed in \cite{Julia-Diaz:2003gq} via algebraic parametrisations of the ground-state wave function.  They also depend on the form of the current-operators, which in \cite{Julia-Diaz:2003gq} are generated by the dynamics from single-quark currents that are Poincar\'e covariant under a kinematic subgroup: i.e., instant-form, point-form \cite{Boffi:2001zb} or front-form \cite{Miller:2002qb,Frank:1995pv,Brodsky:2003pw,Cardarelli:2000tk}.  (NB.\ The point-form construction of spectator-current operators is not fully constrained by Poincar\'{e} invariance alone \cite{Melde:2004qu}.  One can appeal to time-reversal invariance to further limit the composition, but the result still contains a residual arbitrariness which increases with $Q^2$ \cite{Melde:2006}.)  Notably, the studies in \cite{Boffi:2001zb,Cardarelli:2000tk} are predictive and falsifiable because they are based on a mass operator from which wave-functions are actually calculated.  

Instant-form kinematics emphasises features that are closest to those of nonrelativistic quark models, with an interpretation of the wave-function that stresses covariance under three-dimensional rotations and translations.  In models based on light-front kinematics the relevant Lorentz boosts and translations are all kinematic, whereas in the point-form approach there are no kinematical translations and also no kinematic interpretation of the wave function as a representation of spatial structure.  In a relativistic approach the addition of spins to form the total angular momentum of the composite system always requires momentum-dependent rotations of constituent spins \cite{Chung:1988my}.  These are the Wigner rotations (related to Melosh rotations in the front-from), which generate a purely kinematical contribution to constituent-quark orbital angular momentum and guarantee that the bound-state wave-function is an eigenfunction of $J$ and $J_z$ in its rest frame.

\begin{figure}[t]

\centerline{
\includegraphics[width=0.66\textwidth,angle=0]{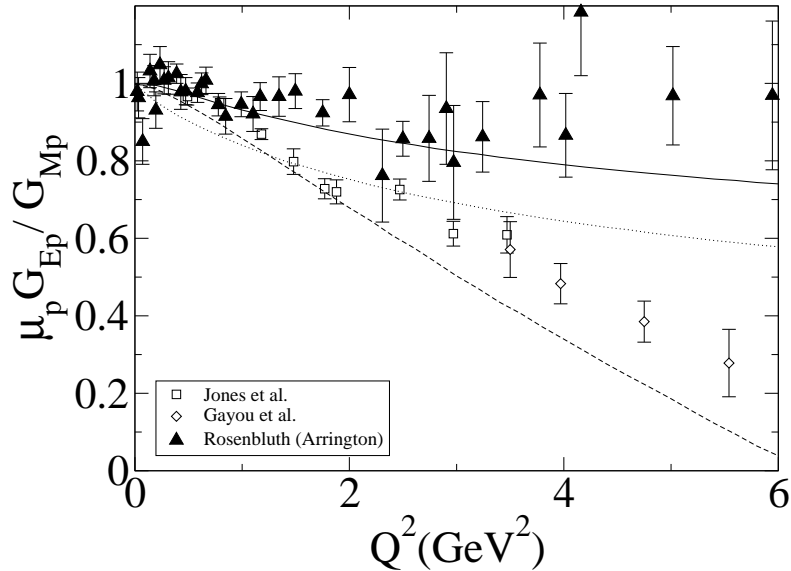}}

\caption{\label{plotGEpGMpCoester} $\mugegmp$ vs.\ $Q^2$ in relativistic quantum mechanics: Solid-curve -- instant form; dotted-curve -- point form; dashed-curve -- front form.  For context, the open symbols indicate the ratio inferred from JLab polarisation transfer data and the filled triangles are from Rosenbluth separation. 
(Figure adapted from \protect\cite{Julia-Diaz:2003gq}.)}
\end{figure}

With an aim of providing a broad perspective, \cite{Julia-Diaz:2003gq} illustrated that, independent of the form of kinematics used in constructing the current operator, a simple algebraic representation of the ground state orbital wave function can yield a fair description of individual elastic form factors.  The proton's electric form factor was found most sensitive to the form of kinematics and the corresponding spectator current.  The best-fit parameters that characterise the wave-function vary markedly from one form of kinematics to another.  These parameters determine the range and shape of the orbital wave-function.  In order to fit the data, instant- and front-form kinematics demand a spatially extended wave function, whereas a compact wave-function is required in point-form kinematics.  To quantify, using one measure of the rms matter radius of the constituent-quark wave-function, the instant- and front-form wave-functions correspond to $r_0 \approx 0.6\,$fm, whereas the point-form wave-function is characterised by $r_0 \approx 0.2\,$fm.  In connection with Figure\;\ref{fig:gep}, and as illustrated in Figure\;\ref{plotGEpGMpCoester}, instant-form dynamics as implemented in \cite{Julia-Diaz:2003gq} gives a ratio that exhibits modest suppression, matching most closely the form factor data inferred via Rosenbluth separation in the one-photon exchange approximation.  The point-form calculation drops more rapidly with increasing $Q^2$; and the front-form result drops most rapidly, lying below but tracking the polarisation transfer data.  It is thus plain that accurate form factor data can be used to test constituent-quark model mass operators.  Equally, however, such mass operators are model-specific and do not possess a veracious connection with QCD dynamics.

\subsection{Three-body problem in quantum field theory}
In quantum field theory a meson (quark-antiquark bound-state) appears as a pole in a four-point quark-antiquark Green function (see, e.g., \cite{Llewellyn-Smith:1969az}, and \cite{Maris:2003vk} for phenomenological applications).  The residue of that pole is proportional to the meson's Bethe-Salpeter amplitude, which is determined by an homogeneous Bethe-Salpeter equation.  It is plain by analogy that a nucleon (three-quark bound-state) must appear as a pole in a six-point quark Green function. (These facts underpin the extraction of masses and form factors through numerical simulations of lattice-regularised QCD, Section~\protect\ref{sec:lattice}.)  In this case the residue of the pole is proportional to the nucleon's Faddeev amplitude, which is obtained from a Poincar\'e covariant Faddeev equation that adds-up all the possible quantum field theoretical exchanges and interactions that can take place between the three dressed-quarks that constitute the nucleon.  

While this is true in principle, the tractable treatment of the Faddeev equation in quantum field theory requires a truncation.  One such scheme is founded \cite{Cahill:1988dx,Reinhardt:1989rw} on the observation that an interaction which describes colour-singlet mesons also generates quark-quark (diquark) correlations in the colour-$\bar 3$ (antitriplet) channel \cite{Cahill:1987qr}.  Naturally, diquarks are confined; namely, they are not asymptotic states in the strong interaction spectrum \cite{Bender:1996bb,Hellstern:1997nv,Bender:2002as,Bhagwat:2004hn,Alkofer:2005ug}.  Reference~\cite{Burden:1988dt} reported a rudimentary study of this Faddeev equation and subsequently more sophisticated analyses have appeared; e.g., \cite{Asami:1995xq,Oettel:1998bk,Hecht:2002ej,Rezaeian:2004nf}.  It has become apparent that the dominant correlations for ground state octet and decuplet baryons are scalar and axial-vector diquarks.  This may be understood on the grounds that: the associated mass-scales are smaller than the masses of these baryons \cite{Burden:1996nh,Maris:2002yu}, with models typically giving masses (in GeV)
\begin{equation}
\label{diquarkmass}
m_{[ud]_{0^+}} = 0.74 - 0.82 \,,\; m_{(uu)_{1^+}}=m_{(ud)_{1^+}}=m_{(dd)_{1^+}}=0.95 - 1.02\,;
\end{equation}
the electromagnetic size of these correlations is less than that of the proton \cite{Maris:2004bp}
\begin{equation}
\label{disize}
r_{[ud]_{0^+}} \approx 0.7\,{\rm fm}\,,
\end{equation}
from which one may estimate $r_{(ud)_{1^+}} \sim 0.8\,{\rm fm}$ based on the $\rho$-meson/$\pi$-meson radius-ratio \cite{Burden:1995ve,Hawes:1998bz}; and the positive parity of the correlations matches that of the baryons.  Both scalar and axial-vector diquarks provide attraction in the Faddeev equation; e.g., a scalar diquark alone provides for a bound octet baryon and including axial-vector correlations reduces that baryon's mass.  

For some it may be noteworthy that the possibility of diquark correlation has been studied in numerical simulations of lattice-regularised quenched-QCD.  Reference \cite{Hess:1998sd} obtains so-called weakly bound scalar and axial-vector diquarks whose masses are in accord with (\ref{diquarkmass}).  This is confirmed in subsequent analyses that are not uniformly restricted to the quenched-truncation; e.g., \cite{Alexandrou:2006cq,Liu:2006zi}.$^\dagger$\footnotetext{It is curious that some refer to the $0^+$ correlation as the \emph{good} diquark and the $1^+$ correlation as the \emph{bad} diquark.  We emphasise that the axial-vector diquark enables dynamical correlations within a baryon that are essential.  A realistic description is forfeited if the axial-vector diquark is omitted.}  In addition, a lattice estimate of diquark size \cite{Alexandrou:2006cq} is consistent with (\ref{disize}).  

The truncation of the Faddeev equation's kernel is completed by specifying that the quarks within the baryon are dressed, with two of the three dressed-quarks correlated always as a colour-$\bar 3$ diquark.  Binding is then effected by the iterated exchange of roles between the bystander and diquark-participant quarks.  This ensures that the Faddeev amplitude exhibits the correct symmetry properties under fermion interchange.  A Ward-Takahashi-identity-preserving electromagnetic current for the baryon thus constituted is subsequently derived~\cite{Oettel:1999gc}.  It depends on the electromagnetic properties of the axial-vector diquark correlation: its magnetic and quadrupole moments, $\mu_{1^+}$ and $\chi_{1^+}$, respectively; and the strength of electromagnetically induced $0^+ \leftrightarrow 1^+$ diquark transitions, $\kappa_{\cal T}$.

Thus does one arrive at an analogue in quantum field theory of the mass- and current-operators necessary in the quantum mechanical treatments described above.  A merit of the Poincar\'e covariant Faddeev equation is that a modern understanding of the structure of dressed-quarks and -gluons is straightforwardly incorporated; viz., effects owing to and arising from the strong momentum dependence of these propagators are realised and exhibited.  This momentum-dependence, which explains, e.g., the connection between constituent- and current-quark masses, was predicted by Dyson-Schwinger equation studies and has been confirmed in lattice simulations.  A synopsis can be found in Section~5.1 of \cite{Holl:2006ni}.

\begin{figure}[t]

\centerline{
\includegraphics[width=0.66\textwidth]{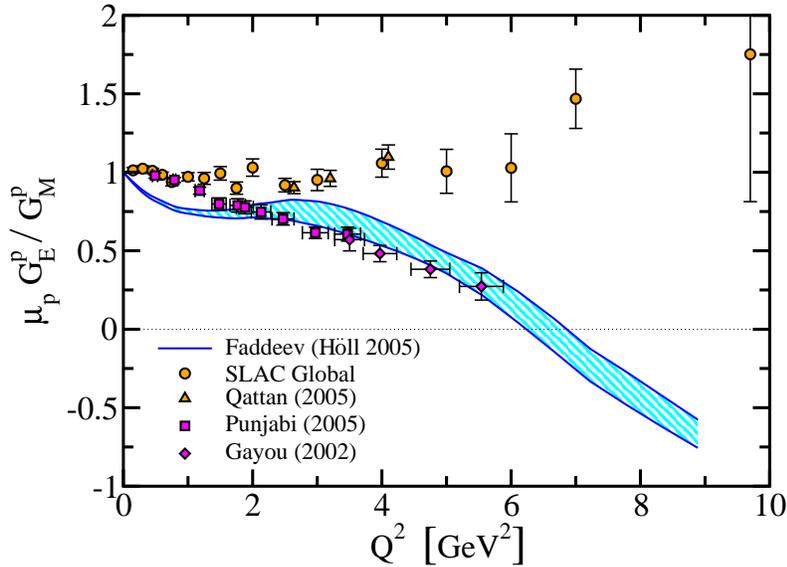}}

\caption{\label{plotGEpGMp} $\mugegmp$ vs.\ $Q^2$.  $\gep$ was calculated using the point-particle values: $\mu_{1^+}=2$ and $\chi_{1^+}=1$, and $\kappa_{\cal T} = 2$.  Variations in the axial-vector diquark parameters used to evaluate $\gep$ have little effect on the results.  The width of the band reflects the variation in $\gmp$ with axial-vector diquark parameters: the upper border is obtained with $\mu_{1^+}=3$, $\chi_{1^+}=1$ and $\kappa_{\cal T}= 2$, while the lower has $\mu_{1^+}= 1$.  
Data: \textit{squares}--\protect\cite{punjabi05}; \textit{diamonds}--\protect\cite{gayou02}; \textit{triangles}--\protect\cite{qattan05} and \textit{circles}--\protect\cite{walker94}.
(Figure adapted from \protect\cite{Holl:2005zi} and prepared with the assistance of A.~H\"oll.)}
\end{figure}

Figure~\ref{plotGEpGMp} depicts the calculated \cite{Holl:2005zi,Alkofer:2004yf} ratio of the proton's Sachs form factors; viz., $\mu_p \gep(Q^2)/\gmp(Q^2)$.   It is important to understand the behaviour of the data cf.\ the calculation at small $Q^2$.  In the neighbourhood of $Q^2=0$, 
\begin{equation}
\mu_p\,\frac{ \gep(Q^2)}{\gmp(Q^2)} = 1 - \frac{Q^2}{6} \,\left[ (r_p)^2 - (r_p^\mu)^2 \right]\,,
\end{equation}
where $r_p$ and $r_p^\mu$ are, respectively, the electric and magnetic radii.  Experimentally, $r_p\approx r_p^\mu$ and this explains why the data varies by less than 10\% on $0<Q^2< 0.6\,$GeV$^2$.  The calculated curve was obtained ignoring the contribution from pion loops, which interfere constructively with those from the axial-vector diquark correlations.  Without such chiral corrections, $r_p> r_p^\mu$ and hence the calculated ratio falls immediately with increasing $Q^2$.  Incorporating pion loops one readily finds $r_p\approx r_p^\mu$ \cite{Alkofer:2004yf}.  It is thus apparent that the small $Q^2$ behaviour of this ratio is materially affected by the proton's pion cloud.  Moreover, such contributions may actually play a role to $Q^2\nlsim 2\,$GeV$^2$.  

These features of pseudoscalar meson contributions to form factors are evident, e.g., in \cite{Miller:2002ig,Alkofer:1993gu,Sato:2000jf,Hammer:2003qv} and we judge that a pointwise accurate description of the individual proton and neutron form factors at small $Q^2$ is impossible without a careful treatment of meson cloud effects.  This is particularly true of the neutron's charge form factor, since $\gen(Q^2=0)=0$ and the form factor is never large, so its evolution is sensitive to delicate cancellations between the quark-core and meson cloud.  Note, however, that merely perceiving a small-$Q^2$ deviation between the form factors and a global dipole fit is not an unambiguous signal of pion cloud effects because \textit{a priori} there is no reason to expect the form factors to be accurately described by a single dipole parametrisation valid uniformly on a large $Q^2$-domain.

Pseudoscalar mesons are not pointlike and hence pion cloud contributions to form factors diminish in magnitude with increasing $Q^2$.  It follows that the evolution of $\mu_p\, \gep(Q^2)/\gmp(Q^2)$ on $Q^2\ngsim 2\,$GeV$^2$ is primarily determined by the quark core of the proton.  This is evident in Figure\;\ref{plotGEpGMp}, which illustrates that for $Q^2 \ngsim 2 \,$GeV$^2$, $\mu_p\, \gep/\gmp$ is sensitive to the parameters defining the axial-vector-diquark--photon vertex.  The ratio passes through zero at $Q^2 \approx 6.5\,$GeV$^2$; namely, at the point for which $\gep(Q^2)=0$.  In this approach the existence of the zero is robust but its location depends on the model's parameters.  The behaviour of $\mu_p\, \gep/\gmp$ owes itself primarily to spin-isospin correlations in the nucleon's Faddeev amplitude.  The forthcoming JLab experiment~\cite{e04108} will test these predictions.  

Naturally, the ratio $\mu_n\,\gen(Q^2)/\gmn(Q^2)$ is also of experimental and theoretical importance.  Notably, in the neighbourhood of $Q^2=0$, 
\begin{equation}
\label{smallQ}
\mu_n\,\frac{ G_E^n(Q^2)}{G_M^n(Q^2)} = - \frac{r_n^2}{6}\, Q^2 ,
\end{equation}
where $r_n$ is the neutron's charge radius.  The Faddeev approach shows  (\ref{smallQ}) to be a good approximation for $r_n^2 Q^2 \nlsim 1$ \cite{Bhagwat:2006py} and extant data \cite{madey03} are consistent with this.  It is thus evident that, as for the proton, this ratio's small $Q^2$ behaviour is materially affected by the neutron's pion cloud.  Reference \cite{Bhagwat:2006py} predicts that the ratio will continue to increase steadily until $Q^2\simeq 8\,$GeV$^2$.  That will be examined in future experiments, Section\;\ref{futureexperiment}.

In common with relativistic constituent-quark models, the Faddeev equation analysis described above makes an assumption about the dynamical content of QCD.  The assumptions can be wrong but they are testable.  Moreover, in this case the Schwinger functions at which one arrives can in principle; viz., at some future time, be calculated via numerical simulations of lattice-regularised QCD.

\begin{figure}[t]

\centerline{
\includegraphics[width=0.66\textwidth,angle=0]{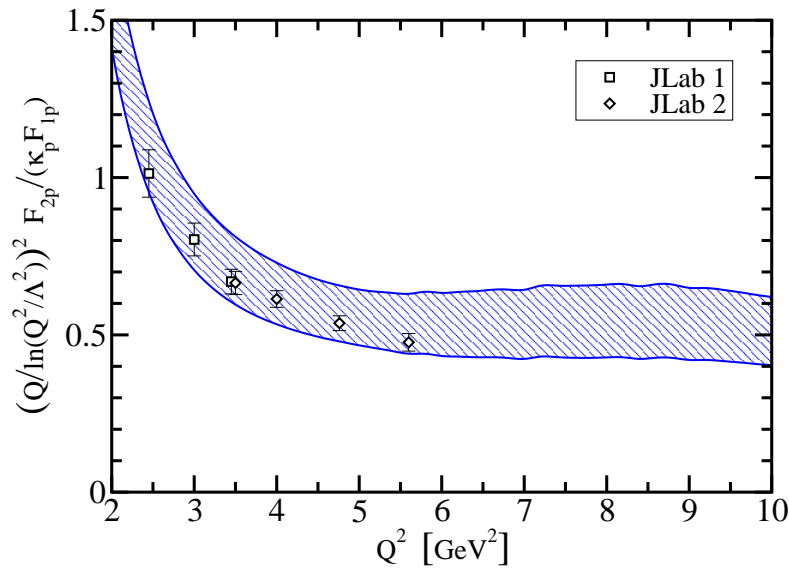}}

\caption{\label{plotF2F1log}  Proton Pauli$/$Dirac form factor ratio, calculated using a Poincar\'e covariant Faddeev equation model \protect\cite{Alkofer:2004yf} and expressed via (\ref{scaling}) with $\Lambda= 0.94\,$GeV.  The band is as described in Figure\;\protect\ref{plotGEpGMp} except that here the upper border is obtained with $\mu_{1^+}=1$, $\chi_{1^+}=1$ and $\kappa_{\cal T}= 2$, and the lower with $\mu_{1^+}=3$.
The data are: \textit{squares}--\protect\cite{punjabi05} and \textit{diamonds}--\protect\cite{gayou02}.
(Figure adapted from \protect\cite{Holl:2005zi} and prepared with the assistance of A.\,H\"oll.)}
\end{figure}

The predictions of perturbative QCD were revisited in an analysis \cite{Belitsky:2002kj} that considers effects arising from both the proton's leading- and subleading-twist light-cone wave functions, the latter of which represents quarks with one unit of orbital angular momentum, with the result
\begin{equation}
\label{scaling}
\frac{Q^2}{[\ln Q^2/\Lambda^2]^{2+\frac{8}{9\beta}}} \, \frac{F_2(Q^2)}{F_1(Q^2)} =\,{\rm constant,}\;\; Q^2\gg \Lambda^2\,,
\end{equation}  
where $\beta=11- \frac{2}{3} N_f$, with $N_f$ the number of quark flavours of mass $\ll Q^2$, and $\Lambda$ is a mass-scale that corresponds to an upper-bound on the domain of nonperturbative (soft) momenta.  This is naturally just a refined version of (\ref{F2F1scale}).  Equation~(\ref{scaling}) is not predictive unless the value of $\Lambda$ is known \textit{a priori}.  However, $\Lambda$ cannot be computed in perturbation theory.  Notwithstanding this the empirical observation was made \cite{Belitsky:2002kj} that on the domain \mbox{$2 \nlsim  Q^2 ({\rm GeV}^2) \nlsim 6$} the $Q^2$-dependence of the polarisation transfer data is approximately described by (\ref{scaling}) with $0.2 \leq \Lambda ({\rm GeV}) \leq 0.4$.  However, that is merely accidental: $\Lambda \simeq 0.3\,$GeV corresponds to a length scale $r_\Lambda \sim 1\,$fm and it is not credible that perturbative QCD is valid at ranges greater than the proton's radius.  In fact, one can argue \cite{Alkofer:2004yf} that a judicious estimate of the least-upper-bound on the domain of soft momenta is $\Lambda = M_N$; viz., the nucleon mass, in which case the Dirac and Pauli form factors obey (\ref{scaling}) for $Q^2\ngsim 5\,$GeV$^2$ as illustrated in Figure\,\ref{plotF2F1log}.

The result in Figure~\ref{plotF2F1log} is not significantly influenced by details of the diquarks' electromagnetic properties.  Instead, the behaviour is primarily governed by correlations expressed in the proton's Faddeev amplitude and, in particular, by the amount of intrinsic quark orbital angular momentum \cite{Bloch:2003vn}.  The nature of the kernel in the Faddeev equation (or, analogously, a mass operator) specifies just how much quark orbital angular momentum is present in a baryon's rest frame.  

It is noteworthy that orbital angular momentum is not a Poincar\'e invariant.  However, if absent in a particular frame, it will inevitably appear in another frame related via a Poincar\'e transformation.  (Therefore is Wigner rotation necessary in constituent-quark models.)  Nonzero quark orbital angular momentum is a necessary outcome of a Poincar\'e covariant description, which is why the covariant Faddeev amplitude is a matrix-valued function with a rich structure that, in a baryons' rest frame, corresponds to a relativistic wave function with $s$-, $p$- and even $d$-wave components.  This is well illustrated by Figure\;6 in \cite{Oettel:1998bk}, which explicitly depicts these components of the Faddeev wave function in the nucleon's rest frame.  A crude estimate based on their magnitudes indicates that the probability for a $u$-quark to carry the proton's spin is $P_{u\uparrow}\sim 80\,$\%, with $P_{u\downarrow}\sim 5\,$\%, $P_{d\uparrow}\sim 5\,$\% and $P_{d\downarrow}\sim 10\,$\%.  Hence, by this reckoning $\sim 30$\% of the proton's rest-frame spin is located in dressed-quark angular momentum.

Elastic electromagnetic form factors can be expressed as one-dimensional integrals of valence-quark generalised parton distributions (GPDs), which are nonforward matrix elements of light-front operators.  It follows that accurate experimental measurements and theoretical calculations of electromagnetic form factors can be used to place much-needed stringent constraints on parametrisations of GPDs.  With such parametrisations in hand one can, e.g., estimate the contribution of quark spin and orbital angular momentum to the light-front nucleon spin \cite{Afanasev:1999at,Burkardt:2004bv,Guidal:2004nd,Wakamatsu:2006dy}.

\begin{figure}[t]

\centerline{
\includegraphics[width=0.66\textwidth,angle=0]{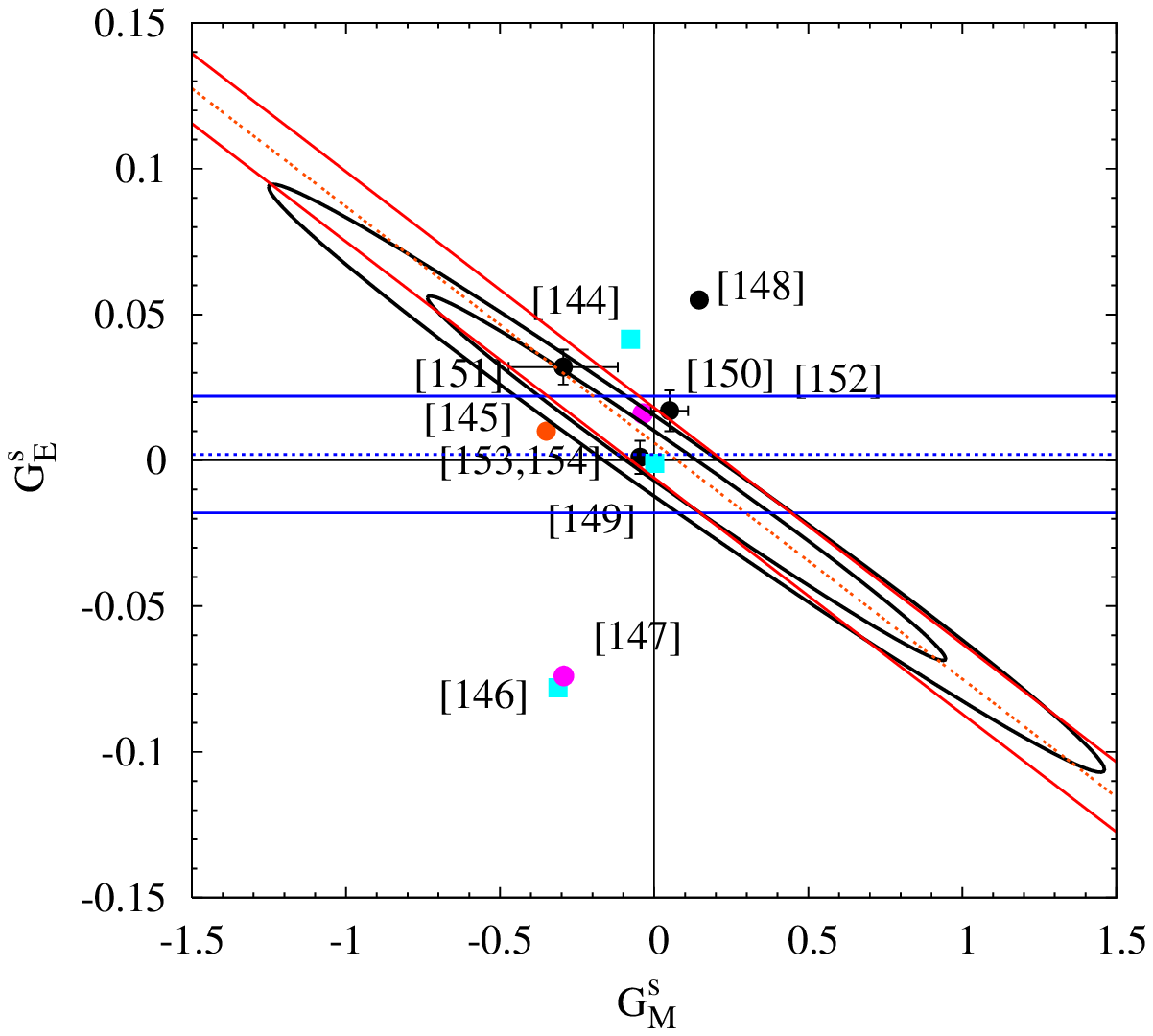}}

\caption{The contours display the 68 and 95\% confidence intervals from an analysis \cite{Young:2006jc} of $G^{ps}_E$ and $G^{ps}_M$ at $Q^2=0.1\,$GeV$^2$, together with estimates of these quantities from model calculations
\cite{Park:1990is,Musolf:1993fu,Forkel:1995ff,Hammer:1995de,Weigel:1995jc,%
Meissner:1997qt,Lyubovitskij:2002ng} and numerical simulations of lattice-regularised QCD \cite{Dong:1997xr,Lewis:2002ix,Leinweber:2004tc,Leinweber:2006ug}.  Recent JLab measurements \cite{acha06} not included in \cite{Young:2006jc} are overlaid for comparison: \textit{diagonal band} -- HAPPEx-H; \textit{horizontal band} -- HAPPEx-$^4\!$He.  (Figure composed with assistance of R.\,D.~Young.)
\label{fig:0604010}}
\end{figure}

\subsection{Strangeness in the proton}
\label{strangep}
As we described briefly in Sections~\ref{sec:ffintro} and \ref{expflavour}, the $s$-quark contribution to the proton's form factors is accessible via parity violating electron scattering if one has accurately determined $G_{E,M}^p(Q^2)$, $G_{E,M}^n(Q^2)$.  Naturally, since the nucleon has no net strangeness, $G_E^{ps}(0)=0$.  However, there is no such simple constraint on either the sign or magnitude of $\mu_s^p=G_M^{ps}(0)$.  In analogy with (\ref{nradius}), a strangeness charge-radius can be defined via
\begin{equation}
\label{rps}
\langle r_{ps}^2\rangle = -6 \left.\frac{dG_E^{ps}(Q^2)}{d Q^2}\right|_{Q^2=0}.
\end{equation} 

In Figure\;\ref{fig:0604010} we provide a snapshot of the current status of experiment and theory for the strange form factors of the proton.  
One model estimate lies within the $95$\% confidence limit \cite{Meissner:1997qt}.  It is inferred from a dispersion-relation fit to nucleon electromagnetic form factors and yields $\mu_s^p= 0.003$, $r_{ps}^2= 0.002\,$fm$^2$.
%
%
%
The lattice-QCD estimates are described in Section\;\ref{sec:lattice}.  
It has been argued \cite{Young:2006jc} that at present the world's data are consistent with the strange form factors of the proton being zero.  Forthcoming experimental results from JLab and Mainz will contribute more to the picture that is developing of the form factors accessible through parity violating electron scattering.

\subsection{Two-photon exchange: Rosenbluth and polarisation transfer}\label{sec:twophoton}
It is apparent from our discussion that theory views the polarisation transfer data as the truest measure of $\gegmp$.  There is a mounting body of evidence that the discrepancy between this data and that obtained via Rosenbluth separation arises from the effects of two-photon exchange (TPE) contributions to the cross-section, which are only partially accounted for in the standard treatment of radiative corrections \cite{mo69}.  For example, it was demonstrated \cite{guichon03} that TPE contributions could explain the discrepancy without spoiling the linear $\varepsilon$-dependence of the reduced cross section.  Moreover, improved evaluations of the effect of the exchange of an additional soft photon \cite{maximon00,arrington04c} had already indicated that the effects were larger than previously believed.
\begin{figure}[t]

\centerline{
\includegraphics[width=0.45\textwidth,angle=90]{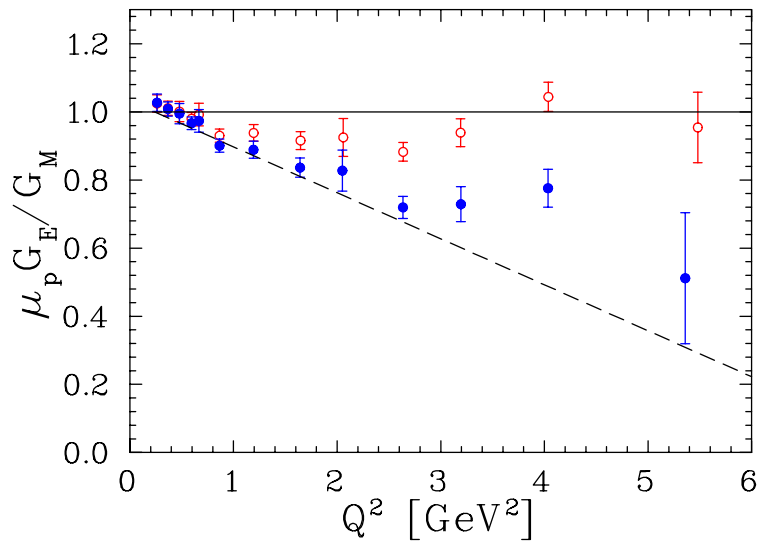}}

\caption{New version of a global analysis of the L-T (Rosenbluth) data on $\mugegm$ \cite{arrington04a}.  Open circles -- ratio obtained from cross-sections \emph{before} two-photon correction; closed circles -- the same ratio obtained after correcting the cross-sections for two-photon exchange \cite{blunden05a}.  Dashed line: a fit to the high-$Q^2$ polarisation transfer data \cite{arrington04a}.
\label{fig:e01001}}
\end{figure}

Figure\;\ref{fig:e01001} illustrates the impact that TPE corrections can have. Calculations including hard and soft photons have recently been performed in both hadronic \cite{blunden05a,kondratyuk05,jain06,borisyuk06} and partonic \cite{afanasev05a} models.  However, at this time there is no complete calculation, valid at all relevant kinematics.  Hadronic calculations generally account only for an unexcited proton in the hadronic intermediate state.  They are thus limited
to relatively low $Q^2$ where the contributions of the excited intermediate
states should be small.  Contributions beyond the elastic, including the
$\Delta$ \cite{kondratyuk05} and higher mass states are under investigation.  The partonic calculations are only valid at large $s$, $t$ and $Q^2$ values,
and rely on a model of the proton's GPDs.  One can also make empirical estimates of the TPE amplitudes \cite{guichon03,arrington04d} but this requires assumptions about the angular dependence of the amplitudes and is limited by the precision of the Rosenbluth data when using the discrepancy as a measure of the size of the TPE effects.  

These calculations and empirical estimates of the TPE amplitudes make several common predictions.  First, all can at least partially resolve the discrepancy between the Rosenbluth and polarisation transfer results, with TPE mainly changing $\gep$ as extracted from Rosenbluth separations.  They all point to relatively weak $Q^2$-dependence, meaning that the corrections could also impact upon precision measurements at low $Q^2$.  Finally, they indicate that the effects are largest at large scattering angles, corresponding to $\varepsilon \rightarrow 0$, and small for $\varepsilon \rightarrow 1$, which implies that there will also be an impact on the extraction of $\gmp$ from the extrapolation of $\sigma_R$ to $\varepsilon=0$, (\ref{eq:sigr}).

The idea that TPE corrections could be large enough to explain the discrepancy seemed originally to contradict limits set by previous measurements designed to test the Born approximation.  
First, the linear $\varepsilon$-dependence of the reduced cross-section is consistent with one-photon exchange and
second, several comparisons of $e^+$--$p$ and $e^-$--$p$ (and $\mu^\pm$--$p$) scattering  showed no indication of TPE effects (see, e.g., \cite{mo69,mar68,camilleri69,drell59, campbell69, greenhut69}).
The interference between Born and TPE amplitudes changes sign when the sign of the charge of one of the scattered particles is flipped, and these measurements indicated that TPE effects were small, on the order of 1\%, while the discrepancy implies missing corrections of up to 6\% \cite{arrington04a}. 
Finally, measurements of the normal-polarisation in polarisation transfer experiments \cite{bizot65,giorgio65, lundquist68} and the asymmetry, $A_N$, in beam-target asymmetry measurements \cite{chen68, powell70}, are consistent with zero, in accordance with the Born approximation expectation.

It has since been found that each of these earlier measurements had insufficient sensitivity or kinematic coverage to observe TPE effects of the scale predicted by contemporary calculations.  Owing to low luminosity of the secondary positron and muon beams, almost all the comparisons were at very low $Q^2$, or at small scattering angles, corresponding to $\varepsilon>0.7$.  The limited data at smaller $\varepsilon$, on the domain where calculations indicate that TPE contributions are largest, provide evidence for a TPE correction of $(4.9 \pm 1.4)$\% \cite{arrington04b}.  Similarly, while a global analysis was able to set tight limits on the size of nonlinearities in the $\varepsilon$-dependence of the reduced cross section \cite{tvaskis06}, these limits are consistent with the calculated nonlinearities \cite{blunden05a, afanasev05a}.  New positron measurements will be performed in the near future to better measure the effect of TPE at large angles and moderate $Q^2$ values \cite{e04116, vepp_proposal}.  Further high-precision Rosenbluth separation measurements \cite{e05017}, using the proton detection technique of  \cite{qattan05}, will provide the sensitivity needed to observe the calculated nonlinearities.  It will also provide clean measurements of TPE effects on the cross-section at large $Q^2$ values, where the contribution of $\gep$ becomes negligible.

At present the form factor discrepancy provides indirect evidence for TPE, while the existing positron measurements at large scattering angle provide only a $3\sigma$ indication.  While future measurements will significantly improve the situation, there is one observable that has provided clear evidence for the presence of TPE corrections in elastic $e$--$p$ scattering.  The asymmetry in the scattering of transversely polarised electrons from unpolarised protons, $A_\perp$, is zero in the Born approximation but can be non-zero owing to TPE contributions.  Measurements have shown significant asymmetries in measurements at MIT-Bates \cite{wells01} and Mainz \cite{maas05}.  Both $A_\perp$ and $A_N$ are sensitive to the imaginary part of the TPE amplitude, while the form factors depend on the real part of the amplitude.  Hence these measurements do not provide \textit{direct} quantitative information that can be applied to the form factor measurements.  However, they do provide a clean signal for the presence of TPE and can be used to test calculations of the TPE amplitudes.

There is an ongoing experimental and theoretical program aimed at fully understanding TPE corrections.  
Whereas much of the focus has been on the discrepancy in $\gep$ at large $Q^2$, TPE corrections are also important at low $Q^2$ and in extracting $\gmp$.  Since TPE corrections vary slowly with $Q^2$ their effects will also be important for high-precision measurements at low $Q^2$.  While the \textit{fractional} TPE correction to $\gep$ is large, it is typically comparable to or only slightly larger than the experimental uncertainties.  The correction to $\gmp$ is much smaller, typically at the few percent level, but this is larger than the quoted experimental uncertainties.  Hence the correction to $\gmp$ can potentially have a \textit{larger} impact in constraining calculations of the form factors.

The quantitative study of the effects of TPE has only just begun.  The impact of these corrections at low $Q^2$ has been evaluated in extractions of the proton charge radius \cite{blunden05b} and the $s$-quark contribution to the nucleon form factors \cite{arrington06b}.  However, these form factors are important input for many analyses.  The effects of TPE corrections must therefore be more quantitatively understood so that we may have the reliable input needed to study; e.g., hyperfine splitting in hydrogen \cite{brodsky04b}, L-T separations in quasielastic scattering \cite{dutta03}, extractions of the axial form factor from neutrino scattering \cite{budd03}, and VCS/DVCS measurements \cite{hydewright04} that require precise knowledge of the Bethe-Heitler process in order to extract the Compton scattering amplitudes.  It is not always sufficient to provide corrected form factors, as many of these processes will also be modified by TPE and the corrections will not be identical for all observables. 

\section{Lattice QCD}
\label{sec:lattice}
Section\;\ref{experimentJA} explains that nucleon form factors have been studied experimentally to very high precision, while Section\;\ref{theoryCDR} exhibits that this precision data is available on the nonperturbative domain of QCD whereupon models and truncations of QCD have been extensively applied.  On this domain the numerical simulation of lattice-regularised QCD, when used in conjunction with chiral effective field theory, is becoming an important addition to the array of tools available in modern hadron physics.  Moreover, it is widely hoped that at some future time this approach may yield uniquely reliable predictions for hadron observables from QCD.  
%
%


\subsection{Overview}
The lattice study of hadron electromagnetic form factors has a long history. The first calculations of the pion form factor were performed over 20 years ago, initially with $SU(2)$ colour \cite{Wilcox:1985iy,Woloshyn:1985in,Woloshyn:1985vd} and subsequently $SU(3)$ colour \cite{Draper:1988bp,Martinelli:1987bh}.  They were promptly followed by calculations of the proton's electric form factor \cite{Martinelli:1988rr}.  Soon thereafter, magnetic moments and electric charge radii were extracted from the $\pi,\ \rho$ and $N$ electromagnetic form factors \cite{Draper:1989pi}.  Despite the limited computing resources of the late '80s and early '90s, it proved possible to perform calculations of the electromagnetic properties, including magnetic moments and charge radii, of the entire baryon octet \cite{Leinweber:1990dv} and decuplet \cite{Leinweber:1992hy}, with results that could be compared with experiment.  The first attempt at examining the $Q^2$-dependence of nucleon electromagnetic form factors was reported in \cite{Wilcox:1991cq}, where the authors found positive values for $\gen$.

There has recently been renewed interest in calculating electromagnetic form factors on the lattice.  For example, the QCDSF collaboration performed a quenched-QCD analysis of these form factors at momentum transfers $Q^2 \in (0.45,1.95)\,$GeV$^2$ \cite{Capitani:1998ff}.  (NB.\ The domain is fixed and restricted by the lattice regularisation.  According to Section\;\ref{sec:ffintro}, such momentum transfers resolve length-scales $d \sim 0.14 - 0.3\,$fm.  The proton's charge radius is $r_p =0.85\,$fm.)  The study was performed at three quark masses and three lattice spacings, allowing the chiral and continuum limits to be investigated.  Upon comparison with experiment, the authors observed the lattice dipole masses to be too large [see the discussion of (\ref{latticeFit})] and attributed this to quenching.  However, the limited spatial resolution might also bear some of the responsibility.

This analysis was extended in \cite{Gockeler:2003ay} to include an extraction of magnetic moments and charge radii.  That allowed for a comparison with chiral effective field theory (ChEFT).  The authors derived an extrapolating function from ChEFT that was then compared with the lattice results, which were calculated at heavy quark masses ($m_\pi>500$ MeV).  Their extrapolations of the proton and neutron anomalous magnetic moments confirmed predictions
\cite{Leinweber:1992hj,Leinweber:1998ej,Hackett-Jones:2000qk,Hackett-Jones:2000js,Leinweber:2001ui,Hemmert:2002uh}
that substantial curvature is required between the lowest-$m_\pi$ lattice-data point and the chiral limit in order to obtain agreement between the lattice results and experiment.  This substantial curvature was also predicted in \cite{Young:2004tb},
where quenched lattice results were extrapolated to the chiral limit using finite range regularisation \cite{Leinweber:2003dg}.

Hitherto the most extensive study of $\gen$ was carried out in quenched-QCD \cite{Tang:2003jh} with $Q^2\in(0.3, 1.0)\,$GeV$^2$. (This corresponds to a resolution of length-scales $d \sim 0.20$ -- $0.36\,$fm.  The scale associated with the neutron radius is 0.58\,fm.)  The authors observed a positive form factor, to which a Galster parametrisation \cite{galster71} was fitted and therefrom a value inferred for the neutron charge radius at each of the simulated quark masses.  Various methods of extrapolating to the chiral limit were subsequently considered.  It was argued that the radius deduced could be reconciled with experiment by standard phenomenology and lowest- or next-to-lowest-order contributions from chiral perturbation theory.  A result obtained more directly is lacking.

The renewed focus on form factor calculations may be attributed to the challenge of new precision data, and improvements in both algorithms and machine speed, which are enabling simulations to be performed with dynamical quarks, at quark masses corresponding to $m_\pi\gtorder$~300~GeV, on lattices with spacing $a \sim 0.1\,$fm and spatial extent $L\sim 2.5\,$fm.  This brings the promise of more realistic simulation results.  It is nonetheless sobering that the computational demands are estimated to increase as $(1/m_\pi)^9$.


\subsection{Magnetic Moments and Form Factor Radii}
\label{MagMom}
As explained in association with (\ref{genrad}), radii are extracted from a form factor's slope at $Q^2=0$, and magnetic moments are obtained from $G_M(Q^2=0)$ [see the discussion associated with (\ref{GEpeq})].  However, lattice kinematics entails that $Q^2=0$ is not directly accessible in the simulations.  Hence an extrapolation from $Q^2\neq 0$ is required in order to infer values of these static properties and one commonly fits the form factor results using a dipole:
\begin{equation}
\label{latticeFit}
F(Q^2) = \frac{F(0)}{(1+Q^2/m_D^2)^2}\ ,
\end{equation}
with $F(0)$ the fitted normalisation and $m_D$ the fitted dipole mass.

A large statistics investigation of the electromagnetic properties of the octet baryons in quenched-QCD has recently been performed \cite{Boinepalli:2006xd}.  Magnetic moments, and electric and magnetic radii were extracted from the form factors for each individual quark-flavour in order to test the environmental sensitivity of the quark contributions to these quantities.  Simulations were performed with pion masses as low as 300\,MeV in order to search for evidence of the chiral nonanalytic behavior predicted by quenched chiral perturbation theory.  Of particular interest was an observed environmental isospin dependence of the strange quark distributions in $\Lambda^0$ and $\Sigma^0$.  It was found that when the environmental quarks are in an isospin-0 state ($\Lambda$ baryon) the strange quark distribution is broader than when the environmental quarks are in an isospin-1 state ($\Sigma$ baryon).

An up-to-date study of isovector nucleon electromagnetic form factors, (\ref{isovectorF}), at momentum transfers $Q^2\in (0.2,2.5)\,$GeV$^2$ is reported in \cite{Alexandrou:2006ru}.  This kinematic domain probes length-scales in the range $(0.12,0.4)\,$fm.  The calculations were performed in quenched-QCD and in $N_f=2$-QCD.  At the pion masses accessible in this study; i.e., $m_\pi \ngsim 0.4\,$GeV, the effects of unquenching were perceived to be small -- an observation consistent with the analysis of \cite{Young:2004tb}.  In comparison with experiment, the lattice results for the isovector form factors in \cite{Alexandrou:2006ru} again lie uniformly above the data.  For the ratio $R_p$ described in Section\;\ref{sub:back}, the lattice results are constant out to $Q^2=2.5\,$GeV$^2$ whereas the polarisation transfer data show suppression (see Figure\;\ref{fig:gep}).  With a chiral extrapolation used to estimate the isovector magnetic moment in the chiral limit, agreement with the experimental value was obtained.  On the other hand, consistent with earlier calculations, the charge radius is constant on the domain of quark masses employed and hence disagrees with experiment.  It is possible that the mismatch with experiment owes to current-quark masses that are too far from reality and/or finite lattice spacing effects.  Reference\;\cite{Alexandrou:2006ru} noted that if dynamical fermions and small quark masses are needed, then a realistic study will require large computer resources.    

\begin{figure}[t]
\begin{tabular}{lr}
\includegraphics[width=8cm]{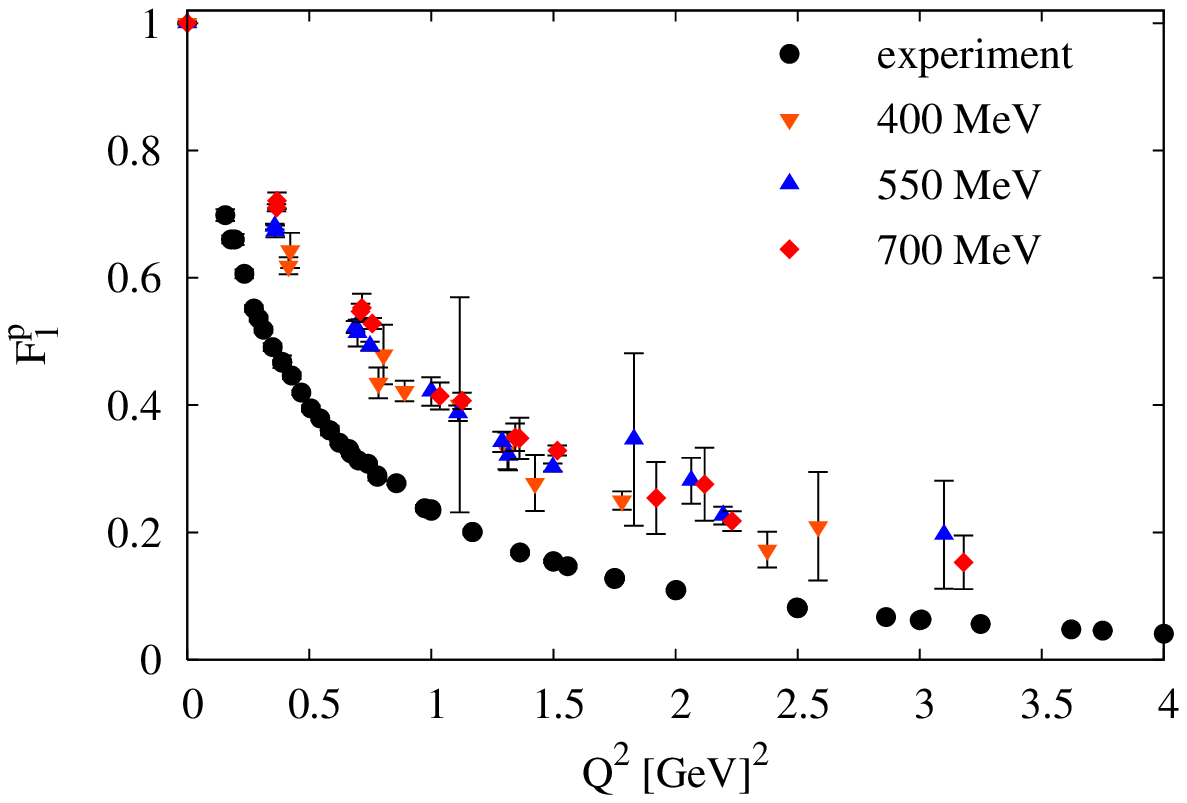}
&
\hspace*{-5mm}
\includegraphics[width=8cm]{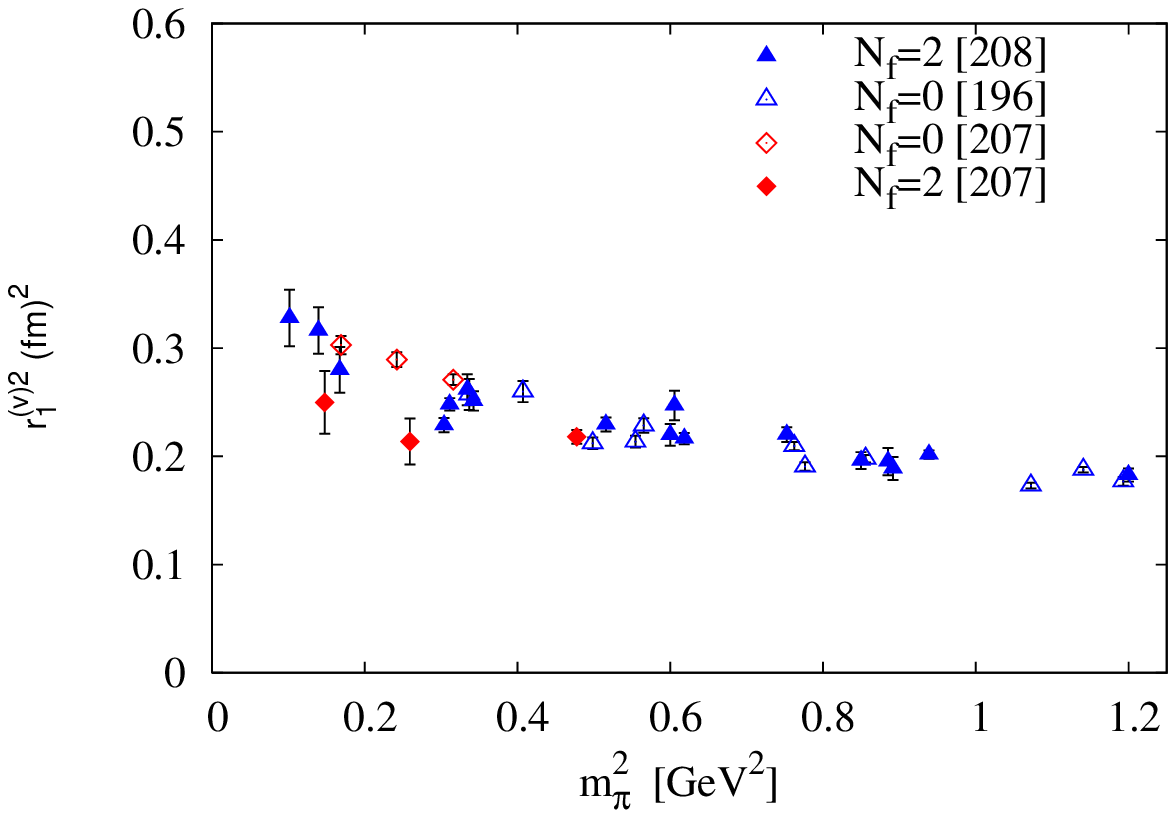}
\end{tabular}
\caption{\label{fig:f1} Left panel: $F_1$ form factor for three different pion masses compared with experimental data \cite{Gockeler-QCDSF-FF}.  Right panel: Isovector $F_1$ (Dirac) form factor radius as a function of $m_\pi^2$ from both quenched-QCD \cite{Gockeler:2003ay,Alexandrou:2006ru} and $N_f=2$-QCD \cite{Alexandrou:2006ru,Gockeler-QCDSF-FF}.}
\end{figure}

Aspects of the behaviour just described are illustrated in Figure\;\ref{fig:f1}.  The left panel depicts a typical example of the proton's Dirac form factor calculated at three different pion masses \cite{Gockeler-QCDSF-FF}.  The results sit high cf.\ experiment, with a slight trend towards the experimental points as $m_\pi$ is decreased.  This translates into small form factor radii that increase with decreasing pion mass, as seen in the right panel, wherein we display quenched-QCD
\cite{Gockeler:2003ay,Boinepalli:2006xd,Alexandrou:2006ru} and $N_f=2$-QCD \cite{Alexandrou:2006ru,Gockeler-QCDSF-FF} results for the isovector $F_1$ form factor radius, plotted as a function of $m_\pi^2$.  The figure indicates a gradual evolution toward the chiral limit.  However, ChEFT predicts that both the $F_1$ and $F_2$ radii should increase dramatically in the neighbourhood of the chiral limit \cite{Gockeler:2003ay,Young:2004tb}.  In modern studies there are indications of incipient ``chiral curvature'' for $m_\pi \nlsim 400$~MeV.  Recent results from the LHPC Collaboration \cite{Edwards:2006qx} also exhibit the features described here.

\begin{figure}[t]
\begin{tabular}{lr}
\includegraphics[width=8cm]{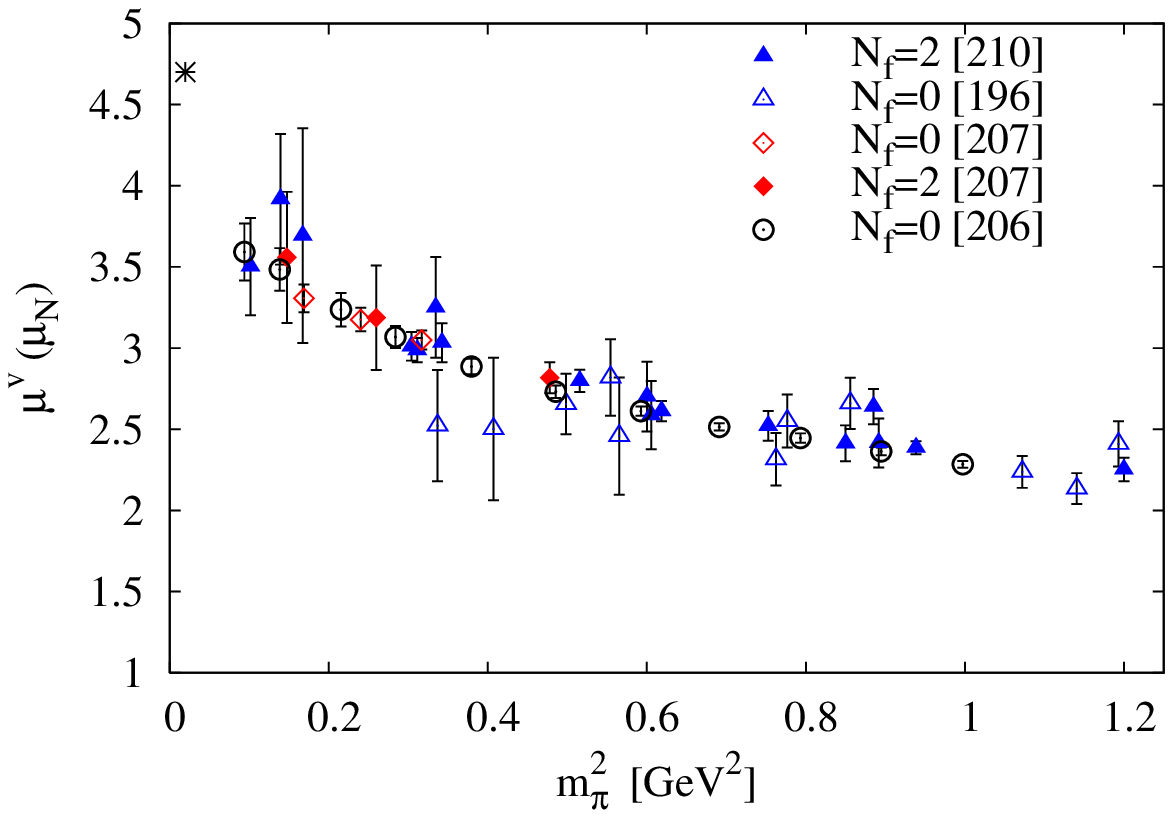}
&
\hspace*{-5mm}
\includegraphics[width=8cm]{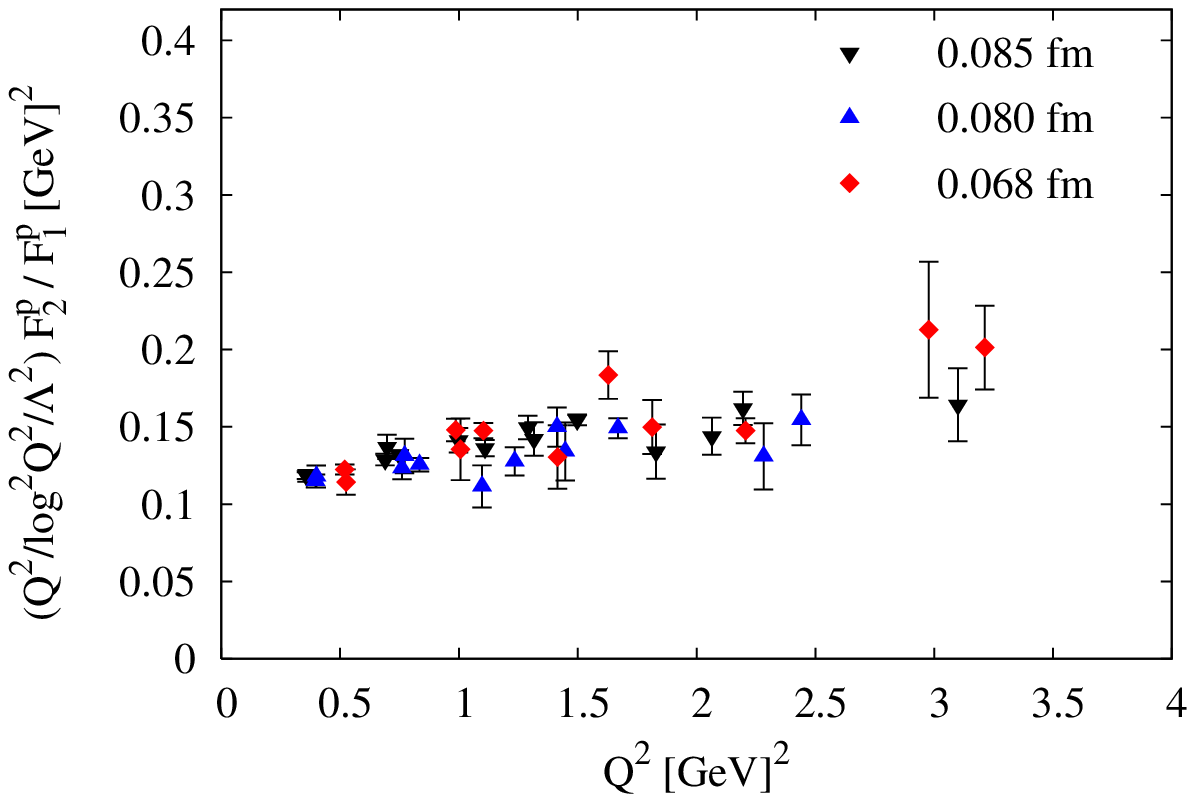}
\end{tabular}
\caption{\label{fig:MagMom} Left panel: Summary of the latest lattice results for the isovector magnetic moment as a function of $m_\pi^2$ from both quenched-QCD \cite{Gockeler:2003ay,Boinepalli:2006xd,Alexandrou:2006ru} and $N_f=2$-QCD \cite{Alexandrou:2006ru,Gockeler:2006ui}.  Right panel: $F_2/F_1$ form factor ratio at $m_\pi\approx 550$~MeV and three different lattice spacings \cite{Gockeler-QCDSF-FF,Gockeler:2006ui} plotted according to (\protect\ref{scaling}) with $\Lambda=0.2\,$GeV.}
\end{figure}

The left panel of Figure\;\ref{fig:MagMom} depicts the latest lattice results for the isovector magnetic moment (inferred by fitting (\ref{latticeFit}) on the accessible kinematic domain and listed in units of lattice nuclear magnetons, $\mu_N=e/2M^{\rm latt}$, where $M^{\rm latt}$ is the nucleon mass measured in the lattice simulation) as a function of $m_\pi^2$, obtained using both quenched-QCD \cite{Gockeler:2003ay,Boinepalli:2006xd,Alexandrou:2006ru} and $N_f=2$-QCD \cite{Alexandrou:2006ru,Gockeler:2006ui}.  The experimental value is indicated by a star at the physical pion mass.  Consistent with \cite{Young:2004tb}, there appears to be little difference between the results obtained in quenched and unquenched simulations.  The evolution with $m_\pi^2$ appears gentle.  However, it is plain that a linear extrapolation would miss the experimental point by many standard deviations.  This, too, is compatible with chiral perturbation theory, which predicts a rapid increase in $\mu_N$ as $m_\pi \to 0$ \cite{Gockeler:2003ay,Young:2004tb}.  (There is a hint of this chiral curvature at the smaller $m_\pi$ values in Figure\;\ref{fig:MagMom}.)


\subsection{Investigating the $Q^2$ dependence and reaching for large $Q^2$}
%
As reviewed in Section~\ref{sub:back}, counting rules for QCD's hard amplitudes predict $F_1(Q^2) \sim 1/Q^4$ and $F_2(Q^2) \sim 1/Q^6$.  It is currently difficult, however, to obtain lattice results with high enough precision over a large enough range of $Q^2$ values to distinguish between dipole and quadrupole behaviour.  Furthermore, it is extremely likely that the present limitations on lattice kinematics prevent a determination of $\zeta_{\rm pQCD}$ and an exploration of the domain $Q^2> \zeta_{\rm pQCD}^2 \gg \Lambda_{\rm QCD}^2$ on which such behaviour should be apparent.  

Some attempts have nevertheless been made in connection with the ratio $F_2(Q^2)/F_1(Q^2)$, whose perturbative scaling is described by (\ref{scaling}).  The right panel of Figure \ref{fig:MagMom} displays values for the ratio in (\ref{scaling}) calculated with $\Lambda =0.2\,$GeV from results at three lattice spacings with approximately the same pion mass \cite{Gockeler-QCDSF-FF,Gockeler:2006ui}: the ratio is roughly constant on the domain of $Q^2$ explored.  This notwithstanding, as discussed in connection with (\ref{scaling}) and Figure\;\ref{plotF2F1log}, such a low value of $\Lambda$ is not credible as the least-upper-bound on the domain of soft momenta in QCD.

On the qualitative side, the results in Fig.~\ref{fig:MagMom}, together with those in
\cite{Alexandrou:2006ru}, indicate that lattice spacing, quark mass and quenching effects on $F_2(Q^2)/F_1(Q^2)$ appear small in comparison with statistical errors.  Quantitatively, though, the lattice values are higher than the corresponding experimental data. Thus, while contemporary lattice simulations are able to reproduce qualitative features of the experimental data, smaller pion masses, at least, are needed before a quantitative description can become a reality.


Herein we have focused attention on the $Q^2$-evolution of the ratio $\mu_p\gep(Q^2)/\gmp(Q^2)$ because of the intriguing possibility that it will pass through zero at some $Q^2\ngsim 5\,$GeV$^2$.  In Section\;\ref{MagMom} we highlighted $N_f=2$-QCD studies \cite{Alexandrou:2006ru} that explore nucleon form factors for $Q^2\in (0.2,2.5)\,$GeV$^2$ and remarked that the ratio reported therein is constant (or perhaps grows slightly) with increasing $Q^2$ in marked contrast to JLab's polarisation transfer data, Figure\;\ref{fig:gep}.  There is a hint in the lattice results that with the lightest accessible current-quark mass the ratio dips below one at $Q^2\approx 1.5$~GeV$^2$.  This provides encouragement for lattice simulations planned at more realistic quark masses.

While simulations with lighter quark masses are beginning to become possible, the problem remains that a typical lattice calculation is restricted to a small domain of relatively low momenta.  This has inspired the LHPC Collaboration to assess the computation cost for a calculation of the electromagnetic form factors out to
$Q^2\sim 6$~GeV$^2$ \cite{Edwards:2005kw}.  They focused on the Breit frame, $(\vec{p}\,' = -\vec{p} = 2\pi \vec{n}/L)$, because previous studies indicated that therein the form factors have smaller statistical uncertainties than with other momentum combinations at the same $Q^2$.  Their analysis reveals that the relative error in $F_1^{v}(Q^2)$, (\ref{isovectorF}), at fixed pion mass increases as $n^4$.  Since their point with $n^2=4$ ($Q^2=4.15$~GeV$^2$) has a relative error of 62\%, then in order to achieve a point at $n^2=8$ ($Q^2\approx 6$~GeV$^2$) with a relative error of 30\%, they would have to increase the statistical accuracy by at least a factor of 50.  Furthermore, to compound the difficulty, it was observed that the relative error in the isovector Dirac form factor increased with approximately the fourth power of $1/m_\pi$.  It is evident therefore that an essentially new technique must be found before numerical simulations of lattice-regularised QCD are in a position to calculate the form factors at large $Q^2$ with realistic quark masses.

\subsection{Strangeness content of the nucleon}
The determination of the strange quark content of the nucleon offers a unique opportunity to obtain information on the role of hidden flavour in the structure of the nucleon (see Sections\;\ref{sec:ffintro}, \ref{expflavour} and \ref{strangep}).

Direct lattice-QCD calculations of the strangeness content are computationally demanding in the extreme and results have so far proved to be inconclusive \cite{Dong:1997xr,Lewis:2002ix,Mathur:2000cf}.  While there is optimism that increases in computing power may enable the next generation of lattice calculations to obtain accurate results, this may actually require an investigation into new lattice techniques.  One possibility is the background field method
\cite{Bernard:1982yu,Martinelli:1982cb,Burkardt:1996vb,Detmold:2004kw}, where a weak signal may be enhanced by coupling a strong electromagnetic field to the vacuum strange quarks.  This technique has recently been employed successfully to calculate physical quantities, such as magnetic moments \cite{Lee:2005ds}, and electric \cite{Christensen:2004ca} and magnetic \cite{Lee:2005dq} polarisabilities.  Alternatively, a method of evaluating the all-to-all propagator, developed by the Dublin group \cite{Foley:2005ac}, offers significantly improved precision over traditional stochastic estimators, and it would be interesting to see this employed in a strangeness form factor calculation.  

Meanwhile, one must continue to rely on more indirect methods for an extraction of the strangeness form factors.  Constraints of charge symmetry were combined with chiral extrapolation techniques, based on finite-range-regularisation, and low-mass quenched-QCD lattice simulations of the individual quark contributions to the charge radii and magnetic moments of the nucleon octet, to obtain precise estimates of the proton's strange electric charge radius \cite{Leinweber:2006ug} and magnetic moment \cite{Leinweber:2004tc}:
\begin{equation}
\langle r^2\rangle^p_s = +0.001\pm 0.004 \pm 0.002\ {\rm fm}^2\,,\;
\mu_s = -0.046\pm 0.019\ \mu_N\,.
\end{equation}
Together, these results considerably constrain the role of hidden flavour in the structure of the nucleon.  Moreover, Figure \ref{fig:0604010} shows that they agree extremely well with a recent analysis \cite{Young:2006jc} of the world's complete data set on parity violating electron scattering.  They are also consistent with the latest measurements from Jefferson Lab~\cite{acha06}, which are indicated by the coloured bands in the figure. 

\section{Epilogue}
\label{Epilepsy}
The world's hadron physics facilities are providing data of unprecedented accuracy, from both nuclear and hadronic targets.  Herein we have focused on nucleon elastic electromagnetic form factors but, before closing, a short digression on the pion is worthwhile.  In that case, too, the impact on our understanding of the basic features of QCD is enormous.  

The pion is notionally a two-body bound-state and hence the simplest composite system described by QCD.  However, the pion is also QCD's Goldstone mode and its properties are intimately connected with confinement and dynamical chiral symmetry breaking (DCSB).  This dichotomy can only be reconciled, and a veracious explanation of pion properties thereby achieved, through a treatment using the full machinery of quantum field theory, and the approach employed in this must guarantee that the axial-vector Ward-Takahashi identity is accurately realised \cite{Maris:1997hd}.  DCSB is expressed through this identity, and there are strong indications that DCSB is a necessary consequence of confinement.  Thus, information on the pion probes into the deepest part of QCD.  New data \cite{Horn:2006tm} and the reanalysis of old data \cite{Tadevosyan:2006yd} are confronting theory, e.g. \cite{Nesterenko:1982gc,Maris:2000sk}.  Through this, reaching to higher momentum transfers promises to identify those elements fundamental to a precise understanding.  Moreover, there is room for some optimism that the JLab upgrade will enable data to be acquired on a domain \cite{Maris:1998hc} in which there are hints of the transition to truly perturbative behaviour.

Returning to nucleon elastic electromagnetic form factors, the next few years will see new neutron data and an extension of proton electric form factor data to $Q^2=8.5\,$GeV$^2$, while the middle of the next decade should see precision neutron and proton data to $Q^2 \approx 14\,$GeV$^2$.  This is part of a much larger programme that will yield an enormous body of concrete information about the spectrum of hadrons and the interactions between them.  An accurate understanding of this data will draw a map of the distribution of mass and spin within the nucleon; lay out the connection between the current-quark and the constituent-quark; and should enable us to become certain of the domain on which perturbative QCD becomes predictive.  

\ack
This work was supported by: Department of Energy, Office of Nuclear Physics, contract no.\ DE-AC02-06CH11357; and PPARC grant PP/D000238/1.   The authors thank those of their colleagues who assisted in the preparation of this overview and apologise for omissions made necessary by the constraints of length and time.

\section{References}
\bibliographystyle{unsrt_with_abbreviations}
\bibliography{form_factors_jpg}

\pagebreak
\tableofcontents

\end{document}